\documentclass[prd,eqsecnum]{revtex4}
\usepackage{graphicx}
\usepackage{dcolumn}
\usepackage{bm}
\newcommand{\be}{\begin{equation}}
\newcommand{\ee}{\end{equation}}
\newcommand{\bse}{\begin{subequations}}
\newcommand{\ese}{\end{subequations}}
\newcommand{\bary}{\begin{eqnarray}}
\newcommand{\eary}{\end{eqnarray}}

\begin{document}

\preprint{ICN/000-03-HEP}

\title{Propagation of Neutrinos through Magnetized Gamma-Ray Burst Fireball}

\author{ Sarira Sahu$^1$\footnote{ Email address: sarira@nucleares.unam.mx }
, Nissim Fraija$^1$\footnote{Email address: nissim.ilich@nucleares.unam.mx }
and Yong-Yeon Keum$^{2,3}$\footnote{Email address: yykeum@korea.ac.kr}
 }
 \affiliation{$^1$Instituto de Ciencias Nucleares,
Universidad Nacional Aut\'onoma de M\'exico,\\
Circuito Exterior, Ciudad Universitaria, Apartado Postal 70-543, 04510 M\'exico DF,
M\'exico\\
$^{2}$Department of Physics and BK21 Initiative for Global Leaders in Physics,
Korea University, Seoul 136-701, Korea \\
$^3$School of Physics, Korea Institute for Advanced Study, Seoul 130-012, Korea
}

\begin{abstract}

The neutrino self-energy is calculated in a weakly magnetized plasma consists of
electrons, protons, neutrons and their anti-particles and using this we have calculated
the neutrino effective potential up to order $M^{-4}_W$. 
In the absence
of magnetic field it reduces to the
known result. We have also calculated explicitly the effective potentials for
different backgrounds which may be helpful in different environments. By considering
the mixing of three active neutrinos in the medium with the magnetic field we have
derived the survival and conversion probabilities of neutrinos from one flavor to
another and also the resonance condition is derived. As an application of the above,
we considered the dense and relativistic plasma of the Gamma-Ray Bursts fireball
through which neutrinos of 5-30 MeV can propagate and depending on the fireball parameters
they may oscillate resonantly or non-resonantly from one flavor to another. These
MeV neutrinos are produced due to stellar
collapse or merger events which trigger the Gamma-Ray Burst. The fireball itself also
produces MeV neutrinos due to electron positron annihilation, inverse beta decay and
nucleonic bremsstrahlung. Using the three neutrino mixing and considering the best
fit values of the neutrino parameters, we found that electron neutrinos are hard to
oscillate to another flavors. On the other hand, the muon neutrinos and the tau neutrinos
oscillate with equal probability to one another, which depends on the neutrino
energy, temperature and size of the fireball. Comparison of oscillation probabilities
with and without magnetic field shows that, they depend on the neutrino energy and also
on the size of the fireball. By using the resonance condition, we have also estimated the
resonance length of the propagating neutrinos as well as the baryon content of the
fireball.

\end{abstract}

\maketitle

\section{Introduction}

The particle propagation in a heat bath with or without magnetic field
has attracted much attention due to its potential importance in
plasma physics, astrophysics and cosmology. The processes which are
forbidden in vacuum can take place in the medium and  even massless
particles acquire mass when they propagate through the medium. Studying
the behavior of particles in such environments requires the technique of
thermal field theory. Therefore, in connection with these astrophysical and
cosmological scenarios it has become increasingly important to
understand the quantum field theory of elementary processes in the
presence of a thermal heat bath. The neutrino self-energy is studied
in the magnetized medium by many authors, where the effective potential of neutrino is
calculated and applied in the physics of supernovae, early Universe and physics of Gamma-ray bursts (GRBs)\cite{Elmfors:1996gy,Koers:2005ya,Dessart:2008zd,Sahu:2009iy,Langacker:1982ih,BravoGarcia:2007uc}.

Gamma-ray bursts (GRBs) are the most luminous objects after the Big Bang in
the universe\cite{Piran:1999kx} and believed to emit about
$10^{51}-10^{55}$ erg in few seconds.
During this few seconds non-thermal flashes of about 100 keV to
1-5 MeV photons are emitted. The isotropic distribution of GRBs\cite{Meegan:1992xg} in
the sky implies that they are of cosmological origin
\cite{Piran:1999kx,Zhang:2003uk,Piran:1999bk}. The GRBs are classified into two
categories: short-hard bursts ($\le 2~s$) and long-soft bursts. It is generally
accepted that long gamma-ray bursts are associated with star forming regions,
more specifically related to supernovae of type Ib and Ic. The observed
correlations of the following GRBs with supernovae GRB 980425/SN 1998bw,
GRB 021211/SN 2002lt, GRB 030329/SN 2003dh and GRB 0131203/SN 2003lw show
that long duration GRBs are related to the core collapse of massive stars\cite{DellaValle:2005cr}.
The origin of short-duration bursts are still a mystery, but recently there
has been tremendous progress   due  to accurate localization of many short
bursts by the Swift\cite{Gehrels:2005qk,Barthelmy:2005bx}
and HETE-2\cite{Villasenor:2005xj} satellites. The afterglow observation of
GRB 050709 at z=0.1606\cite{Hjorth:2005} by HETE-2 and the Swift
observation of afterglow from GRB050709b at z=0.225\cite{Gehrels:2005qk}
and GRB 050724 at z=0.258\cite{Berger:2005} seems to support the
coalescing of compact binaries as the progenitor for the short-hard bursts
although definite conclusions can not be drawn at this stage. Very recently
millisecond magnetars have been considered as possible candidates as
the progenitor for the short-hard bursts\cite{Usov:1992zd,Uzdensky:2007uf}.
For a future study of short-hard GRBs, the ultra-fast flash observatory (UFFO) project is
proposed\cite{uffo:2009}.

Irrespective of the nature of the progenitor or the emission mechanism of the
gamma-rays, these huge energies within a very small volume imply the formation
of $e^{\pm}$ and $\gamma$ fireball which would expand relativistically.
In the standard {\it fireball} scenario; at the first, a radiation dominated
plasma is formed in a compact region with a size $c\delta t\sim 100$-$1000$
km\cite{Piran:1999kx,Waxman:2003vh}. This creates an opaque
$\gamma-e^{\pm}$ fireball due to the process $\gamma+\gamma\rightarrow e^+ +
e^-$. However, in addition to $\gamma$, $e^{\pm}$  pairs, fireball also contain a small
amount of baryons, both from the progenitor and the surrounding medium and the electrons
associated with the matter (baryons),  that increase the opacity and delay the
process of emission of radiation.  The average optical depth of this
process is very high. Because of this huge optical depth\cite{Goodman:1986az}, photons can not
escape freely and even if there are no pairs to
begin with, they will form very rapidly and will Compton scatter lower energy
photons. In the fireball the $\gamma$ and $e^{\pm}$ pairs will thermalize with a
temperature of about 3-10 MeV.

In this stage, a phase  of acceleration begins and the fireball expands
relativistically with a large Lorentz factor, converting internal energy into bulk kinetic
energy. As the fireball shell expands, the baryons will be accelerated by radiation pressure.
The fireball bulk Lorentz factor increases linearly with radius, until reaching the maximum
Lorentz factor, so the photon number density and typical energy drop. At certain
radius, the photons become optically thin (the optical depth is $\tau_{\gamma\gamma}\simeq
1$) to both pair production and to Compton scattering off the free electrons
associated with baryons. At this radius, although much of the initial energy is
converted to the kinetic energy  of the shell, some energy will be radiate away with
an approximately black body spectrum. This is the first electromagnetic signal detectable
from the fireball. For an intermittent central engine with typical variability timescale
of $\delta t$, appears adjacent mini-shells with different Lorentz factor, which will
collide with each other and will form strong "internal" shocks. Later, the fireball shell
is eventually decelerated by successive strong external shocks with the ambient medium (ISM),
propagates into the medium\cite{Zhang:2003uk}. As in each shell exists a non thermal
population of baryons and electrons through Fermi acceleration and during each
shock the system behaves like an inelastic collision between two or more
shells converting kinetic energy into internal energy, which is given to the
non thermal population of baryons and electrons  cool via synchrotron emission
and/or inverse Compton scattering to produce the observed prompt emission.
\cite{Piran:1999kx}.  The synchrotron spectrum can be calculate if we know the
detailed physical conditions of the radiating region. For internal shocks, the
so-called 'equipartition' hypothesis is often used, which assumes that the
energy is equally distributed between protons, electrons and the magnetic
field\cite{Vedrenne}.

As is well known, during the final stage of the death of a massive star and/or
merger of binary stars copious amount of neutrinos in the energy range of 5-30 MeV
are produced. Some of these objects are possible progenitors of GRBs \cite{Lee:2007js}. 
Apart from the beta decay process there many other processes which are responsible for the
production of MeV neutrinos in the above scenarios: for example, electron-positron
annihilation, nucleonic bremsstrahlung etc, where neutrinos of all flavor can be
produced\cite{Raffelt:2001kv}. Within the fireball, inverse beta decay as well as
electron-positron annihilation will also produce MeV neutrinos. Many of these neutrinos
 has been intensively studied in the literature \cite{Ruffert:1998qg,Goodman:1986we}
 and may propagate through the fireball. However, the high-energy neutrinos created
 by photo-meson production of pions in interactions between the fireball gamma-rays
 and accelerated protons have been studied too\cite{Waxman:1997ti}. The accretion disc
 formed during the collapse  or merger is also a potentially important place to produce
 neutrinos of similar energy.  Fractions of these neutrinos will propagate through the
 fireball and they will oscillate \cite{Volkas:1999gb,Dasgupta:2008cu}
if the accreting  materials survive
for a longer period. Although neutrinos conversions in a polarized medium have been
 studied \cite{Nunokawa:1997dp}, resent we have studied
 the neutrino propagation  within the fireball environment with and without magnetic
 field where resonant neutrino  oscillate from one flavor to
another are studied by considering the mixing of two flavors only.
In this paper we have calculated the neutrino effective potential in the weak
field and done a complete analysis of the three neutrino mixing
within the magnetized fireball and studied the resonant oscillation of it. By considering
the best fit neutrino parameters from different experiments, we found that electron neutrino
can hardly oscillate to other flavor, whereas muon and tau neutrinos  can oscillate among
themselves with almost equal probability and their oscillation probabilities depends on
neutrino parameters as well as the fireball parameters.

The organization of the paper is as follows: In sec. 2, we have derived the neutrino
self-energy by using the real time formalism of finite temperature field theory
\cite{Nieves:1990ne,Weldon:1982aq,D'Olivo:2002sp,Erdas:1990gy} and
Schwinger's proper-time method \cite{Schwinger}.
By considering the weak-field approximation
we have derived the neutrino effective potential and
compare it with the effective potential for $B=0$ case. We also
calculate the effective potential for matter background as well as
for neutrino background. A  brief description about the Gamma-Ray Burst and the
fireball model is discussed in sec. 3. The case of three-neutrino mixing is
considered in sec. 4, where we have calculated the survival and conversion
probabilities of neutrinos and also the resonance condition.  In sec. 5, we
discuss our results for GRB fireball and a brief conclusions is drawn in sec. 6.


\section{Neutrino Effective Potential}

As is well known, the particle properties get modified when it is immersed in
a heat bath. A massless neutrino acquires an effective mass and an effective
potential in the
medium. The resonant conversion of neutrino from one flavor to another due to
the medium effect is important for solar neutrinos which is well known as the
MSW effect. Similarly the propagation of neutrino in the early universe hot
plasma\cite{Enqvist:1990ad}, supernova medium\cite{Sahu:1998jh} and in the
GRB fireball\cite{Sahu:2005zh} can have also
many important implications in their respective physics.
In all the astrophysical and cosmological environment, magnetic field is
entangled intrinsically with the matter and it also affect the particle
properties. Although neutrino can not couple directly to the magnetic field,
its effect can be felt through coupling to charge particles in the
background\cite{Erdas:1998uu}.
Neutrino propagation in a neutron star in the presence of a magnetic field and
also in the magnetized plasma of the early universe has been studied
extensively.

We use the field theory formalism to study the effect of heat bath and
magnetic field on the propagation of elementary particles. The effect of
magnetic field is taken into account through Schwinger's propertime method\cite{Schwinger}.
The effective potential of a particle is calculated from the real part of its
self energy diagram.

The most general decomposition of the neutrino-self energy in presence of a
magnetized medium can be written as:
\be
\Sigma (k) = R \Big( a_\parallel {\rlap /k}_\parallel
+ a_\perp {\rlap /k}_\perp + b {\rlap /u} + c {\rlap /b} \Big)L \,,
\label{sigmanb}
\ee
where $k^\mu_\parallel=(k^0, k^3)$, $k^\mu_\perp=(k^1,
k^2)$ and $u^\mu$ stands for the 4-velocity of the center-of-mass of the
medium given by $u^\mu = (1, {\bf 0})$.
The
projection operators are conventionally defined as $R=
\frac12(1+\gamma_5)$ and $L = \frac12(1-\gamma_5)$.
The effect of magnetic field enters through the 4-vector
$b^\mu$ which is given by $b^\mu = (0, {\hat {\bf b}})$.
The background classical magnetic field vector is along
the $z$-axis and consequently $b^\mu=(0,0,0,1)$. So using the four vectors
 $u^\mu$ and  $b^\mu$ we can express
\be
{\rlap /k}_\parallel= k_0 {\rlap /u} - k_3 {\rlap /b},
\label{paralk}
\ee
and the self-energy can be expressed in terms of three independent four-vectors
$k^\mu_\perp$, $u^{\mu}$ and $b^{\mu}$. So we can write ($\Sigma = R {\tilde
  \Sigma} L$)
\be
{\tilde \Sigma}=a_{\perp}\rlap /k_{\perp}+b\rlap /u+c\rlap /b.
\label{sigmanb2}
\ee
The determinant of ${\rlap /k} -{\tilde \Sigma}$, i.e
\be
det[ {\rlap /k} -{\tilde \Sigma}]=0,
\ee
gives the dispersion relation up to leading order in $a$, $b$ and $c$ as:
\be
k_0-|{\bf k}|=b-c\,\cos\phi-a_{\perp}|{\bf k}|\sin^2\phi=V_{eff, B},
\label{poteff}
\ee
for a particle, where $\phi$ is the angle between the neutrino momentum and
the magnetic field vector. One has to remember that the scalars
$b$ and $c$ in this case are not the same if one expresses the self-energy in
the form given in Eq.~(\ref{sigmanb}), but the $V_{eff, B}$ is independent of how
we express $\tilde \Sigma$.
Now the Lorentz scalars $a$, $b$ and $c$ which are functions of neutrino
energy, momentum and magnetic field can be calculated from the neutrino
self-energy due to charge current and neutral current interaction of neutrino
with the background particles.

\subsection{Neutrino self-energy}

The one-loop neutrino self-energy in a magnetized medium is comprised
of three pieces\cite{BravoGarcia:2007uc}, one coming from the
$W$-exchange diagram which we
will call $\Sigma_W (k)$, one from the tadpole diagram which we will
designate by $\Sigma_t (k)$ and one from the $Z$-exchange diagram
which will be denoted by $\Sigma_Z (k)$. The total self-energy of the
neutrino in a magnetized medium then becomes:
\be
\Sigma(k) = \Sigma_W(k) + \Sigma_Z (k)+ \Sigma_t (k)\,.
\label{tsen}
\ee
Each of the individual terms appearing in the right-hand side of the
above equation can be expressed as in Eq.~(\ref{sigmanb}) and the Lorentz
scalars $a$, $b$ and $c$ have contributions from all the three pieces as
described above.The individual terms on the right hand side of Eq.~(\ref{tsen}) can be
explicitly written as:
\be
-i\Sigma_W(k)=\int\frac{d^4 p}{(2\pi)^4}\left(\frac{-ig}{\sqrt{2}}\right)
\gamma_\mu\, L \,iS_{\ell}(p)\left(\frac{-ig}{\sqrt{2}}\right)\gamma_\nu
 \,L\,i W^{\mu \nu}(q)\,,
\label{Wexch}
\ee
\be
-i\Sigma_Z(k)=\int\frac{d^4 p}{(2\pi)^4}\left(\frac{-ig}
{\sqrt{2}\cos\theta_W}\right)
\gamma_\mu\, L \,iS_{\nu_\ell}(p)\left(\frac{-ig}{\sqrt{2}\cos\theta_W}\right)
\gamma_\nu\,L\,i Z^{\mu \nu}(q)\,,
\label{Zexch}
\ee
and
\be
-i\Sigma_t(k)= -\left(\frac{g}{2\cos \theta_W}\right)^2 R\,
\gamma_\mu\,iZ^{\mu \nu}(0)\int\frac{d^4 p}{(2\pi)^4} {\rm Tr}
\left[\gamma_\nu \,(C_V + C_A \gamma_5)\,iS_{\ell}(p)\right]\,.
\label{tad}
\ee
The subscripts in $\Sigma$ correspond to W-exchange, Z-exchange and Tadpole diagrams.
In the above expressions $g$ is the weak coupling constant and
$\theta_W$ is the Weinberg angle and $g$ can be expressed in terms of the
Fermi coupling constant as $\sqrt{2} G_F=g^2/4 M^2_W$. The quantities $C_V$ and $C_A$ are
the vector and axial-vector coupling constants
which come in the neutral-current interaction of
electrons, protons ($p$), neutrons ($n$) and neutrinos with the $Z$
boson.  Their forms are as follows,
\bary
C_V=\left \{\begin{array}
{r@{\quad\quad}l}
-\frac{1}{2}+2\sin^2\theta_W & e\\
\frac{1}{2} & {\nu}\\ \frac{1}{2}-2\sin^2\theta_W & {{p}}\\
-\frac{1}{2} & {{n}}
\end{array}\right.,
\label{cv}
\eary
and
\bary
C_A=\left \{\begin{array}
{r@{\quad\quad}l}
-\frac{1}{2} & {\nu},{{p}}\\
\frac{1}{2} & e,{{n}}
\end{array}\right..
\label{ca}
\eary
Here $W^{\mu \nu}(q)$ and $S_\ell(p)$ stand for the $W$-boson propagator
and charged lepton propagator respectively in
presence of a magnetized plasma. The $Z^{\mu \nu}(q)$ is the $Z$-boson
propagator in vacuum and $S_{\nu_\ell}(p)$ is the neutrino propagator
in a thermal bath of neutrinos.  The form of the charged lepton
propagator in a magnetized medium is given by,
\be
S_\ell (p) = S^0_\ell(p) + S^\beta_\ell (p)\,,
\label{slp}
\ee
where $S^0_\ell(p)$ and $S^\beta_\ell (p)$ are the charged lepton
propagators in presence of an uniform background magnetic field and in
a magnetized medium respectively. In this article we will always
assume that the magnetic field is directed towards the $z$-axis of the
coordinate system. With this choice we have,
\be
i S^0_\ell(p) = \int_0^\infty e^{\phi(p,s)} G(p,s)\,ds\,,
\label{sl0p}
\ee
where,
\be
\phi(p,s) = is(p^2_\parallel - m_\ell^2 - \frac{\tan z}{z} p^2_\perp)\,.
\label{phasel}
\ee
In the above expression
\bary
p^2_\parallel &=& p_0^2 - p_3^2\,,\\
p^2_\perp &=& p_1^2 + p_2^2\,,
\eary
and $z= e{ B}s$ where $e$ is the magnitude of
the electron charge, ${B}$ is the magnitude of the magnetic
field and $m_\ell$ is the mass of the charged lepton. In the above
equation we have not written another contribution to the phase which
is $\epsilon |s|$ where $\epsilon$ is an infinitesimal quantity. This
term renders the $s$ integration convergent. We do not explicitly
write this term but implicitly we assume the existence of it and it
will be written if required.  The above equation can also be written
as:
\be
\phi(p,s) = \psi(p_0) - is[p^2_3 + \frac{\tan z}{z} p^2_\perp]\,,
\label{phaselu}
\ee
where,
\be
\psi(p_0)=is(p_0^2 - m_\ell^2)\,.
\label{psi}
\ee
The other term in Eq.~(\ref{sl0p}) is given as:
\be
G(p,s)=\sec^2 z \left[{\rlap A/} + i {\rlap B/} \gamma_5 +m_\ell(\cos^2 z -
i \Sigma^3 \sin z \cos z)\right]\,,
\label{gps}
\ee
where,
\bary
A_\mu &=& p_\mu -\sin^2 z (p\cdot u\,\, u_\mu - p\cdot b \,\,b_\mu)\,,
\label{A}\\
B_\mu &=& \sin z\cos z (p\cdot u \,\,b_\mu - p\cdot b \,\,u_\mu)\,,
\label{B}
\eary
and
\be
\Sigma^3 = \gamma_5 {\rlap /b} {\rlap /u}\,.
\label{sigm3}
\ee
The second term on the right-hand side of Eq.~(\ref{slp}) denotes the
medium contribution to the charged lepton propagator and its form is
given by:
\be
S^\beta_\ell(p) = i \eta_F(p\cdot u)\int_{-\infty}^\infty e^{\phi(p,s)} G(p,s)
\,ds\,,
\label{slbp}
\ee
where $\eta_F(p\cdot u)$ contains the distribution functions of the
particles in the medium and its form is:
\be
\eta_F (p \cdot u) = \frac{\theta(p\cdot u)}{e^{\beta(p\cdot u -
\mu_\ell)} + 1 } +
\frac{\theta(- p\cdot u)}{e^{-\beta(p\cdot u - \mu_\ell)} + 1}\,,
\label{eta}
\ee
where $\beta$ and $\mu_\ell$ are the inverse of the medium temperature
and the chemical potential of the charged lepton.

The form of the $W$-propagator in presence of a uniform magnetic field
along the $z$-direction is presented in \cite{Erdas:1998uu} and in
this article we only use the linearized (in the magnetic field) form
of it. The reason we assume a linearized form of the $W$-propagator is
because the magnitude of the magnetic field we consider is such that
$eB \ll {M^2_W}$. In this limit  and in unitary gauge the
propagator is given by

\be
W^{\mu \nu}(q) = \frac{g^{\mu \nu}}{M^2_W}\left(1 + \frac{q^2}{M^2_W}\right)
- \frac{q^\mu q^\nu}{M^4_W} + \frac{3ie}{2 M^4_W} F^{\mu \nu}\,,
\label{impw}
\ee
where $M_W$ is the $W$-boson mass.  Here we
assume that $q^2 \ll M_W^2$ and keep terms up to $1/M_W^4$
in the $W$ propagator.

Let us assume that an electron neutrino $\nu_e$ is propagating in the medium
(generalization to other neutrinos is straight forward)
which contain electrons and positrons, protons, neutrons and all types of
neutrinos and anti-neutrinos.

By evaluating the Eq.~(\ref{Wexch}) explicitly we obtain
\be
Re~\Sigma_W (k)=R \biggl [
a_{W\perp} {\rlap /k}_{\perp}+ b_W {\rlap /u} + c_W {\rlap /b}
\biggr ] L,
\ee
where the Lorentz scalars are given by
\bary
a_{W\perp}&=&-\frac{\sqrt2G_F}{M_W^2}\biggl[
\biggl\{E_{\nu_e}(N_e-\bar{N}_e)+ k_3(N_e^0-\bar{N}_e^0)\biggr\}\nonumber\\
&&+\frac{eB}{2\pi^2}\int^\infty_0 dp_3\sum_{n=0}^\infty(2-\delta_{n,0})
\biggl (\frac{m_e^2}{E_n}- \frac{H}{E_n}\biggr)(f_{e,n}+\bar{f}_{e,n})\biggr],
\label{conaw}
\eary
\bary
b_W&=&b_{W0}+{\tilde b}_W\nonumber\\
&=&
\sqrt2 G_F \biggl[\biggl(1+\frac32\frac{m_e^2}{M_W^2}
+\frac{E_{\nu_e}^2}{M_W^2}\biggr)(N_e-\bar{N}_e)+\biggl(\frac{eB}{M_W^2}
+\frac{ E_{\nu_e}k_3}{M_W^2}\biggr)(N_e^0-\bar{N}_e^0)\nonumber\\
&&-\frac{eB}{2\pi^2M_W^2} \int^\infty_0 dp_3
\sum_{n=0}^\infty(2-\delta_{n,0})\biggl\{2k_3E_n\delta_{n,0}
+2E_{\nu_e}\biggl(E_n+\frac{m_e^2}{2E_n}\biggr)\biggr\}(f_{e,n}+\bar{f}_{e,n})\biggr]
\label{conbw}
\eary

and
\bary
c_W&=&c_{W0}+{\tilde c}_W\nonumber\\
&=&\sqrt2
G_F\biggl[\biggl(1+\frac12\frac{m_e^2}{M_W^2}-\frac{k_3^2}{M_W^2}\biggr)(N_e^0-\bar{N}_e^0)
+\biggl(\frac{eB}{M_W^2}-\frac{E_{\nu_e}k_3}{M_W^2}\biggr)(N_e-\bar{N}_e)\nonumber\\
&&-\frac{eB}{2\pi^2M_W^2} \int^\infty_0
dp_3\sum_{n=0}^\infty(2-\delta_{n,0})\biggl\{2E_{\nu_e}
\biggl(E_n-\frac{m_e^2}{2E_n}\biggr)\delta_{n,0}
+2k_3\biggl(E_n-\frac32\frac{m_e^2}{E_n}-
\frac{H}{E_n}\biggr)\biggr\}(f_{e,n}+\bar{f}_{e,n})\biggr].
\label{concw}
\eary
The electron energy in the magnetic field is given by,
\be
E_{e,n}^2=(p_3^2+m_e^2+2neB)=(p_3^2+m_e^2+H).
\ee
From Eqs.~(\ref{conbw}) and (\ref{concw}), we have defined
\be
{\tilde b}_W=
-\sqrt2 G_F\frac{eB}{2\pi^2M_W^2} \int^\infty_0 dp_3
\sum_{n=0}^\infty(2-\delta_{n,0})\biggl\{2k_3E_n\delta_{n,0}+2E_{\nu_e}
\biggl(E_n+\frac{m_e^2}{2E_n}\biggr)\biggr\}(f_{e,n}+\bar{f}_{e,n})\biggr],
\label{tildebw}
\ee
and
\be
{\tilde c}_W=  -\sqrt2 G_F \frac{eB}{2\pi^2M_W^2} \int^\infty_0
dp_3\sum_{n=0}^\infty(2-\delta_{n,0})\biggl\{2E_{\nu_e}
\biggl(E_n-\frac{m_e^2}{2E_n}\biggr)\delta_{n,0}
+2k_3\biggl(E_n-\frac32\frac{m_e^2}{E_n}-
\frac{H}{E_n}\biggr)\biggr\}(f_{e,n}+\bar{f}_{e,n})\biggr].
\label{tildecw}
\ee
In the above equations, the number density of electrons is defined as
\be
N_e=\frac{eB}{2\pi^2}\sum_{n=0}^\infty (2 - \delta_{n,0}) \int_0^\infty dp_3
f_{e,n}
\ee
and the number density of electrons for the Lowest Landau (LL) state which
corresponds to $n=0$ is
\be
N_e^0=\frac{eB}{2\pi^2}\int_0^\infty dp_3 f_{e,0}
\label{ne0}
\ee
We can express the Eq.~(\ref{Zexch}) for Z-exchange as
\be
Re\Sigma_Z(k) =R(a_Z\rlap /k+b_Z \rlap /u)L,
\label{sigz}
\ee
and explicit evaluation gives,
\be
a_Z=\sqrt{2}G_F\biggl[\frac{E_{\nu_e}}{M_Z^2}(N_{\nu_e}-\bar{N}_{\nu_e})
+ \frac23\frac{1}{M_Z^2}\biggl(\langle E_{\nu_e}\rangle N_{\nu_e}
+\langle \bar{E}_{\nu_e}\rangle \bar{N}_{\nu_e}\biggr)\biggr],
\ee
and
\be
b_Z=\sqrt{2}G_F\biggl[(N_{\nu_e}-\bar{N}_{\nu_e})-\frac{8E_\nu}{3M^2_Z}
\biggl(\langle E_{\nu_e}\rangle N_{\nu_e}
+\langle \bar{E}_{\nu_e}\rangle\bar{N}_{\nu_e}\biggr)\biggr].
\ee
In Eq.~(\ref{sigz}) we have a term proportional to $\rlap /k$, because there is no magnetic
field. But using the four vectors $\rlap /u$ and $\rlap /b$ the parallel
component of the four vector $\rlap /k$ can be decomposed as in
Eq.~(\ref{paralk}).
In the calculation of the potential the contribution from these terms will cancel each
other and
only one which will remain is $b_Z$.

From the tadpole diagram Eq.~(\ref{tad}) we get,
\bary
Re\Sigma_t (k) &=&\sqrt2 G_FR\biggl[
\biggl\{C_{V_e}(N_e-\bar{N}_e)+C_{V_p}(N_p-\bar{N}_p)+C_{V_n}(N_n-\bar{N}_n)
+(N_{\nu_e}-\bar{N}_{\nu_e})
\nonumber\\
&&+(N_{\nu_\mu}-\bar{N}_{\nu_\mu})+(N_{\nu_\tau}-\bar{N}_{\nu_\tau})\biggr\} \rlap/u
-C_{A_e} (N_e^0-\bar{N}_e^0)\rlap /b\biggr]L.
\eary
So the different contributions to the neutrino self-energy up to order
$1/M^4_W$ are calculated in a background of $e^+e^-$, nucleons, neutrinos and anti-neutrons.


\subsection{Weak field limit $eB\ll m^2_e$}

In the above subsection, the result obtained is weak compared to the W-boson mass
i.e. $eB\ll M^2_W$. But here we would like to use another limit $eB\ll m^2_e$ that is
magnetic field much weaker compared to the one done in the above subsection. We also
assume that the chemical potential of the background electron gas is much small than the
electron energy ($\mu\ll E_e$). The $\mu=0$ implies CP symmetric medium where number
of electrons equals number of positrons. So by taking $\mu\ll E_e$ we assume that
$N_e > {\bar N}_e$. In a fireball medium this condition can be satisfied because the
excess of electrons will come from the electrons associated with the baryons which will
come from the central engine.

In the weak field limit ($eB\ll m^2_e/e=B_c$) and $\mu\ll E_e$, the electron
distribution function can be written as
\be
f_{e,n}=\frac{1}{ e^{\beta (E_{e,n}-\mu) + 1}} \simeq \sum^{\infty}_{l=0} (-1)^l
  e^{-\beta (E_{e,n}-\mu) (l+1)}.
\ee
Also we shall define
\be
\alpha=\beta\mu(l+1),
\ee
and
\be
\sigma=\beta m_e(l+1).
\ee
Using the above distribution function, the electron number density and other
quantities of interest are given below:
\be
N_e^0-{\bar N}^0_{e}=\frac{1}{\pi^2}\frac{B}{B_c} m^3 \sum_{l=0}^{\infty} (-1)^l\sinh{\alpha}
\,K_1(\sigma)]=\frac{m_e^3}{\pi^2}\left (\frac{B}{B_c}\right ) \Phi_1,
\label{ne02}
\ee
\be
N_e-{\bar N}_{e}=\frac{m^3}{\pi^2} \sum_{l=0}^{\infty} (-1)^l \sinh{\alpha}
\left[
\frac{2}{\sigma} K_2(\sigma)-\frac{B}{B_c} K_1(\sigma)
\right]=\frac{m_e^3}{\pi^2} \Phi_2,
\ee
\be
\frac{eB}{2\pi^2}\int_0^\infty dp_3E_0(f_{e,0}+\bar{f}_{e,0})
=\frac{m_e^4}{\pi^2}\biggl(\frac{B}{B_c}\biggr)
\sum_{l=0}^\infty(-1)^l\cosh\alpha\biggl(K_0(\sigma)+\frac{K_1(\sigma)}{\sigma}\biggl),
\ee
\be
\frac{eB}{2\pi^2}\int_0^\infty dp_3\frac{1}{E_0}(f_{e,0}+\bar{f}_{e,0})
=\frac{m_e^2}{\pi^2}\biggl(\frac{B}{B_c}\biggr)
\sum_{l=0}^\infty(-1)^l\cosh\alpha\,\,K_0(\sigma),
\ee
\bary
\frac{eB}{2\pi^2}\sum_{n=0}^\infty(2-\delta_{n,0})
\int_0^\infty dp_3 E_n(f_{e,n}+\bar{f}_{e,n})
&&=\frac{m_e^2}{\pi^2}\sum_{l=0}^\infty(-1)^l\cosh\alpha\nonumber\\
&&\biggl[\biggl(\frac{6}{\sigma^2}-\frac{B}{B_c}\biggr)K_0(\sigma)
+\biggl(2-\frac{B}{B_c}+\frac{12}{\sigma^2}\biggr)\frac{K_1(\sigma)}{\sigma}\biggr],
\eary
\be
\frac{eB}{2\pi^2}\sum_{n=0}^\infty(2-\delta_{n,0})\int_0^\infty dp_3
\frac{1}{E_n}(f_{e,n}+\bar{f}_{e,n})=\frac{m_e^2}{\pi^2}
\sum_{l=0}^\infty(-1)^l\cosh\alpha\biggl[\frac{2}{\sigma}K_1(\sigma)
-\frac{B}{B_c}K_0(\sigma)\biggr]
\ee
and
\be
\frac{eB}{2\pi^2}\sum_{n=0}^\infty(2-\delta_{n,0})\int_0^\infty dp_3
\frac{H}{E_n}(f_{e,n}+\bar{f}_{e,n})=\frac{m_e^4}{\pi^2}
\sum_{l=0}^\infty(-1)^l\frac{\cosh\alpha}{\sigma^2}\biggl[4K_0(\sigma)
+\frac{8}{\sigma}K_1(\sigma)\biggr].
\ee
All the above quantities are necessary to evaluate the effective potential.


\subsection{Neutrino Potential without Magnetic field}

In the absence of magnetic field the neutrino self-energy and the neutrino
effective potential is calculated earlier\cite{Garcia:2007ij}. In this case
the neutrino self-energy is decomposed as
\be
Re{\tilde\Sigma} (k)= a {\rlap /k}+ b {\rlap /u},
\ee
and the neutrino effective potential for a massless neutrino is given by
\be
V_{eff}= b = \frac{1}{4 E_{\nu}} Tr \left ( {\rlap /k} Re{\tilde\Sigma(k)}\right ).
\ee
By evaluating the right hand side (RHS) up to order $1/M^4_W$ gives,
\bary
V_{eff}&=&\sqrt{2} G_F \biggr[\biggl(1+\frac32 \frac{m^2_e}{M^2_W}\biggr)
(N_e-\bar{N}_e)_{B=0}\nonumber\\
&&-\frac{4}{\pi^2} \left ( \frac{m^2_e}{M_W}\right )^2  E_{\nu_e} \sum_{l=0}^{\infty} (-1)^l \cosh\,\alpha
\biggr \{
\frac{4 K_0(\sigma)}{\sigma^2} + \left (1+ \frac{8}{\sigma^2}\right) \frac{K_1(\sigma)}{\sigma}
\biggl\}
\biggl ].
\label{potenb0}
\eary
Also we have
\be
(N_e-{\bar N}_e)_{B=0}=\frac{m_e^3}{\pi^2} \sum_{l=0}^{\infty} (-1)^l
\sinh{\alpha}\,
\frac{2}{\sigma} K_2(\sigma).
\ee
This is the result obtained in ref.\cite{Garcia:2007ij} up to order $1/M^4_W$ for neutrino
propagating in a medium with only electrons and positrons in it.


\subsection{Comparison of $V_{eff}$ with and without magnetic field}

The neutrino effective potential in a magnetic field is
given in Eq.~(\ref{poteff}). To simplify our calculation we assume that, the
magnetic field is along the direction of the neutrino
propagation so that $\phi=0$ and the $a_{\perp}$ term does not contribute.
Also one has to remember that by taking $B=0$, we should get back the result
obtained in
Eq.~(\ref{potenb0}) and this is only possible when we take $k_3=-E_{\nu}$
in our calculation.
Then the effective potential should be defined as (independent of the
angle $\phi$ is zero or not),
\be
V_{eff,B}=(b-c)/_{k_3=-E_{\nu}}.
\ee
Hence forth we shall replace $k_3$ by $-E_{\nu}$ in our calculation.
This gives
\bary
V_{eff,B} &=&
\sqrt2 G_F \biggl[\biggl(1+\frac32\frac{m_e^2}{M_W^2}
-\frac{eB}{M_W^2}\biggr)(N_e-\bar{N}_e)
-\biggl(1+\frac{m_e^2}{2 M_W^2}-\frac{eB}{M_W^2} \biggr )(N_e^0-\bar{N}_e^0)
\nonumber\\
&&+\frac{eB}{2\pi^2M_W^2} \int^\infty_0 dp_3
\sum_{n=0}^\infty(2-\delta_{n,0})\biggl\{2E_{\nu_e}E_n\delta_{n,0}
-2E_{\nu_e}\biggl(2E_n-\frac{m_e^2}{E_n}
-\frac{H}{E_n}\biggr)\biggr\}(f_{e,n}+\bar{f}_{e,n})\biggr]
\eary
With simplifications this gives,
\bary
V_{eff, B} &=& \sqrt2 G_F \biggl[ \frac{m^3_e}{\pi^2} \sum_{l=0}^{\infty} (-1)^l
\sinh\, \alpha
\biggl \{
\biggl(1+\frac32\frac{m_e^2}{M_W^2}
-\frac{eB}{M_W^2}\biggr)
\biggl(
 \frac{2}{\sigma} K_2(\sigma) - \frac{B}{B_c} K_1(\sigma)
\biggr)\nonumber\\
&&-\frac{B}{B_c}\biggl(1+\frac{m_e^2}{2 M_W^2}-\frac{eB}{M_W^2} \biggr )
K_1(\sigma)
\biggr\}\nonumber\\
&&-\frac{2}{\pi^2}\biggl(\frac{m^2_e}{M_W}\biggr)^2E_{\nu_e} \sum_{l=0}^{\infty} (-1)^l
\cosh\, \alpha
\biggl \{
\biggl(\frac{8}{\sigma^2}-\frac52 \frac{B}{B_c} \biggr )K_0(\sigma)
+\biggl( 2-4 \frac{B}{B_c} + \frac{16}{\sigma^2} \biggr )\frac{K_1(\sigma)}{\sigma}
\biggr\} \biggr].
\label{potenb}
\eary
We can write this in a simpler form as
\be
V_{eff,B}= \sqrt2 G_F \frac{m^3_e}{\pi^2}
\biggl [\Phi_A-\frac{2 m_e E_{\nu}}{M^2_W} \Phi_B \biggr ],
\ee
where the functions $\Phi_A$ and $\Phi_B$ are defined as,
\begin{figure}[t!] 
\vspace{0.5cm}
{\centering
\resizebox*{0.4\textwidth}{0.3\textheight}
{\includegraphics{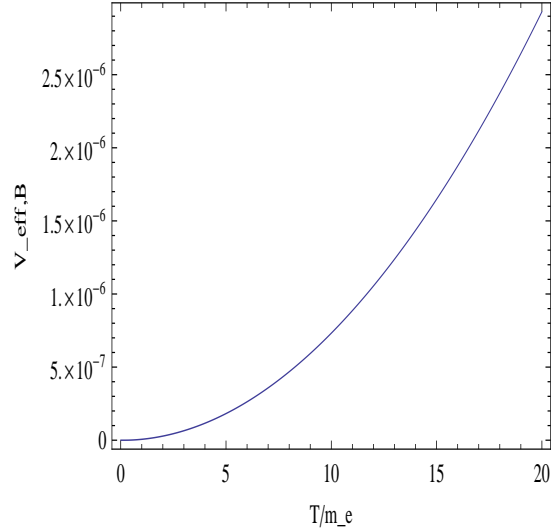}}
}
\caption{\small\sf The Eq.~(\ref{potenb}) is plotted as a function of
temperature $T/m_e$ for a give $B=01.B_c$. The unit of $V_{eff,B}$ is in $eV$.}
\label{figure1}
\end{figure}
\begin{figure}[t!] 
\vspace{0.5cm}
{\centering
\resizebox*{0.4\textwidth}{0.3\textheight}
{\includegraphics{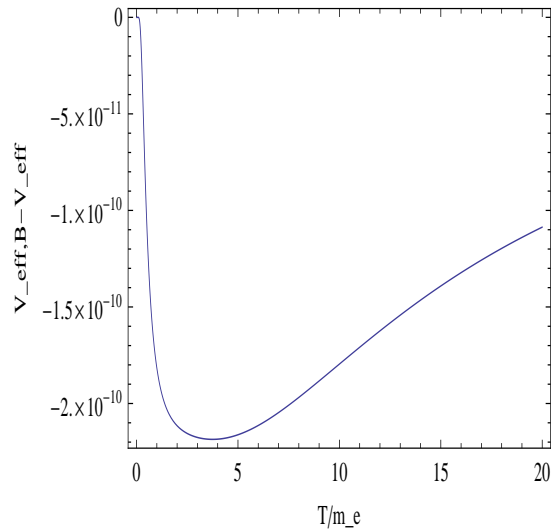}}
}
\caption{\small\sf Only the magnetic field dependence of the $V_{eff,B}$  is
plotted as a function of temperature $T/m_e$ for a given $B=01.B_c$. The unit of
$V_{eff,B}$ is in $eV$.}
\label{figure2}
\end{figure}
%
\bary
\Phi_A&=&\sum_{l=0}^{\infty} (-1)^l
\sinh\, \alpha
\biggl [
\biggl(1+\frac32\frac{m_e^2}{M_W^2}
-\frac{eB}{M_W^2}\biggr)
\biggl(
 \frac{2}{\sigma} K_2(\sigma) - \frac{B}{B_c} K_1(\sigma)
\biggr)\nonumber\\
&&-\frac{B}{B_c}\biggl(1+\frac{m_e^2}{2 M_W^2}-\frac{eB}{M_W^2} \biggr )
K_1(\sigma)
\biggr ],
\label{fphiA}
\eary
and
\be
\Phi_B =
 \sum_{l=0}^{\infty} (-1)^l
\cosh\, \alpha
\biggl [
\biggl(\frac{8}{\sigma^2}-\frac52 \frac{B}{B_c} \biggr )K_0(\sigma)
+\biggl( 2-4 \frac{B}{B_c} + \frac{16}{\sigma^2} \biggr )\frac{K_1(\sigma)}{\sigma}
\biggr ].
\label{fphiB}
\ee
By taking $B=0$ in Eq.~(\ref{potenb}) it reduces to Eq.~(\ref{potenb0}).
So in the weak field limit we get back the potential for $B=0$ in the medium.
Here we have shown only for the W-boson contribution. In Z-exchange diagram
we do not have magnetic field contribution. In the tadpole diagram only
electron loop
will be affected by the magnetic field. But as the momentum transfer is zero,
there will not be any higher order contribution. As the magnetic field is
weak, the
protons and neutrons will not be affected by the magnetic field.

In Fig.~\ref{figure1} we have plotted the potential Eq.~(\ref{potenb}) as
a function of temperature in the range $0$ to 10 MeV for a fixed value of
the magnetic field $B=0.1 B_c$. This shows that
the potential is an increasing function of temperature. We have also shown in
Fig.~\ref{figure2}, only the magnetic field contribution, i.e.
by subtracting the
$B=0$ part from Eq.~(\ref{potenb}), which shows that, the magnetic field
contribution is opposite compared to the medium contribution and also order of
magnitude smaller.


\subsection{Matter Background}

Let us consider the background with electrons, positrons, protons, neutrons,
neutrinos and anti-neutrinos in the background. As we are considering the
magnetic field to be weak, the magnetic field will have no effect on protons
and neutrons. For an  electron neutrino $\nu_e$ propagating in this background, we have
\bary
a_{W\perp}&=&-\frac{\sqrt2G_F}{M_W^2}\biggl[
E_{\nu_e}
\biggl\{(N_e-\bar{N}_e)- (N_e^0-\bar{N}_e^0)\biggr\}\nonumber\\
&&+\frac{m^4_e}{\pi^2}\sum_{l=0}^{\infty} (-1)^l\cosh\alpha
\biggl\{
\left (2-\frac{8}{\sigma^2} \right ) \frac{K_1(\sigma)}{\sigma}
-\left ( \frac{B}{B_c}+\frac{4}{\sigma^2}\right ) K_0(\sigma)
\biggr\}\biggr],
\label{awmatter}
\eary
\bary
b_{e}&=&b_W+b_Z+b_{t}=b_{0e}+\tilde{b}_W\nonumber\\
&=&\sqrt2 G_F \biggl[\biggl(1+\frac32\frac{m_e^2}{M_W^2}
+\frac{E_{\nu_e}^2}{M_W^2}+C_{V_e} \biggr)(N_e-\bar{N}_e)+\biggl(\frac{eB}{M_W^2}
-\frac{ E^2_{\nu_e}}{M_W^2}\biggr)(N_e^0-\bar{N}_e^0)\nonumber\\
&&+C_{V_p}(N_p-\bar{N}_p)+C_{V_n}(N_n-\bar{N}_n)+2(N_{\nu_e}-\bar{N}_{\nu_e})\nonumber\\
&&+(N_{\nu_\mu}-\bar{N}_{\nu_\mu})+(N_{\nu_\tau}-\bar{N}_{\nu_\tau})
-\frac83\frac{E_{\nu_e}}{M_Z^2}\biggl(\langle
E_{\nu_e}\rangle N_{\nu_e}+\langle \bar{E}_{\nu_e}\rangle
\bar{N}_{\nu_e}\biggr)\biggr]+\tilde{b}_W,
\label{bematter}
\eary
and the coefficient of $\rlap /b$ is,
\bary
c_e&=&c_W+c_t=c_{0e}+\tilde{c}_W\nonumber\\
&=&\sqrt{2}G_F\biggl[\biggl(1+\frac{m_e^2}{2M_W^2}-\frac{E^2_{\nu_e}}{M_W^2}
-C_{A_e}\biggr)(N_e^0-\bar{N}_e^0)
+\biggl(\frac{eB}{M_W^2}+\frac{E^2_{\nu_e}}{M_W^2}\biggr)
(N_e-\bar{N}_e)\biggr]+\tilde{c}_W,
\label{cematter}
\eary
where $\tilde{b}_W$ and $\tilde{c}_W$ are given in Eqs.~(\ref{tildebw}) and (\ref{tildecw}).
In the weak field limit these two functions are given as
\be
\tilde{b}_W=-\sqrt{2}G_F\frac{2}{\pi^2}\biggl(\frac{m^2_e}{M_W}\biggr)^2E_{\nu_e} \sum_{l=0}^{\infty} (-1)^l \cosh\, \alpha
\biggl [
\biggl(\frac{6}{\sigma^2}-\frac52 \frac{B}{B_c} \biggr )K_0(\sigma)
+\biggl( 3-2 \frac{B}{B_c} + \frac{12}{\sigma^2} \biggr )\frac{K_1(\sigma)}{\sigma}
\biggr]
\ee
and
\be
\tilde{c}_W=\sqrt{2}G_F\frac{2}{\pi^2}\biggl(\frac{m^2_e}{M_W}\biggr)^2E_{\nu_e} \sum_{l=0}^{\infty} (-1)^l \cosh\, \alpha
\biggl [
\frac{2}{\sigma^2} K_0(\sigma)
-\biggl( 1+2 \frac{B}{B_c} - \frac{4}{\sigma^2} \biggr )\frac{K_1(\sigma)}{\sigma}
\biggr].
\ee

Similarly for muon and tau neutrinos,
\bary
b_{\mu}=b_{0\mu}&=&\sqrt{2}G_F\biggl[C_{V_e}(N_e-\bar{N}_e)
+C_{V_p}(N_p-\bar{N}_p)+C_{V_n}(N_n-\bar{N}_n)+(N_{\nu_e}-\bar{N}_{\nu_e})\nonumber\\
&&+2(N_{\nu_\mu}-\bar{N}_{\nu_\mu})+(N_{\nu_\tau}-\bar{N}_{\nu_\tau})
-\frac83\frac{E_{\nu_\mu}}{M_Z^2}\biggl(\langle
E_{\nu_\mu}\rangle N_{\nu_\mu}+\langle \bar{E}_{\nu_\mu}\rangle \bar{N}_{\nu_\mu}\biggr) \biggr],
\label{bmumatter}
\eary
and
\bary
b_{\tau}=b_{0\tau}&=&\sqrt{2}G_F\biggl[C_{V_e}(N_e-\bar{N}_e)
+C_{V_p}(N_p-\bar{N}_p)+C_{V_n}(N_n-\bar{N}_n)+(N_{\nu_e}-\bar{N}_{\nu_e})\nonumber\\
&&+(N_{\nu_\mu}-\bar{N}_{\nu_\mu})+2(N_{\nu_\tau}-\bar{N}_{\nu_\tau})
-\frac83\frac{E_{\nu_\tau}}{M_Z^2}\biggl(\langle
E_{\nu_\tau}\rangle N_{\nu_\tau}+\langle\bar{E}_{\nu_\tau}\rangle
\bar{N}_{\nu_\tau}\biggr)
  \biggr].
\label{btaumatter}
\eary
respectively and
\bary
c_{\mu} &=& c_{0\mu}=-C_{A_e}(N_e^0-\bar{N}_e^0)\nonumber\\
c_{\tau} &=& c_{0\tau}=-C_{A_e}(N_e^0-\bar{N}_e^0).
\eary
For muon and tau neutrinos propagating in the medium
$\tilde{c}_{\mu}=\tilde{c}_{\tau}=0$.
The matter potentials experience by different neutrinos for $\phi=0$ are given as
\bary
V_{\nu_e}&=&b_e-c_e\nonumber\\
V_{\nu_\mu}&=&b_{\mu} -C_{A_e}(N_e^0-\bar{N}_e^0)\nonumber\\
V_{\nu_\tau}&=&b_{\tau} -C_{A_e}(N_e^0-\bar{N}_e^0).
\eary
Putting the values of $b_l$ and $c_l$  (for $l=e,\mu,\tau$) one can calculate
the neutrino potential in the background.
For charge neutral matter we should impose
\be
N_e-\bar{N}_e=N_p-\bar{N}_p,
\ee
and this gives
\be
C_{V_e}(N_e-\bar{N}_e)+C_{V_p}(N_p-\bar{N}_p)=0,
\ee
in Eqs.~(\ref{bematter}), (\ref{bmumatter}) and (\ref{btaumatter}). The
particle asymmetry is related to the lepton or baryon asymmetry through the
relation
\be
L_a=\frac{N_a-\bar{N}_a}{N_{\gamma}},
\ee
where the number density of photon is $N_{\gamma}=\frac{2}{\pi^2}\zeta(3) T^3$.

\subsection{Only Neutrino Background}
\label{neuback}

In a newly born neutron star, the neutrinos are trapped  because the
mean free path of these neutrinos are very short compared to the depth of the
surrounding medium. So slowly these neutrinos will diffuse out of the region
where they are trapped called the neutrino sphere. In the neutrino sphere, the
different neutrinos have different average energy, which are given as\cite{Giunti}:
\bary
\langle E_{\nu_e}\rangle&\simeq&10\,\,MeV\nonumber\\
\langle \bar{E}_{\nu_e}\rangle&\simeq&15\,\,MeV\nonumber\\
\langle \bar{E}_{\nu_x}\rangle=\langle E_{\nu_x}\rangle &\simeq& 20\,\,MeV,
\eary
for $x=\mu, \tau$.
If the medium contains only the
neutrinos and anti-neutrinos of all flavors, then for propagating
$\nu_e$ and $\nu_{\mu}$ we have
\bary
b_e&=&\sqrt{2}G_F\biggl[2(N_{\nu_e}-\bar{N}_{\nu_e})+(N_{\nu_\mu}-\bar{N}_{\nu_\mu})+(N_{\nu_\tau}-\bar{N}_{\nu_\tau})\nonumber\\
&&-\frac83\frac{\langle E_{\nu_e}\rangle}{M_Z^2}\biggl(\langle E_{\nu_e}\rangle N_{\nu_e} +\langle \bar{E}_{\nu_e}\rangle \bar{N}_{\nu_e}\biggr)\biggr],
\eary
and
\bary
b_\mu&=&\sqrt{2}G_F\biggl[(N_{\nu_e}-\bar{N}_{\nu_e})+2(N_{\nu_\mu}-\bar{N}_{\nu_\mu})+(N_{\nu_\tau}-\bar{N}_{\nu_\tau})\nonumber\\
&&-\frac83\frac{1}{M_Z^2}\biggl(\langle E_{\nu_\mu}\rangle^2 N_{\nu_\mu}+\langle \bar{E}_{\nu_\mu}\rangle^2
\bar{N}_{\nu_\mu}\biggr)\biggr],
\label{neubmu}
\eary
respectively and by interchanging $\mu\leftrightarrow \tau$ in
Eq.~(\ref{neubmu}) we obtain $b_{\tau}$ for tau neutrino.
For only neutrino background we have $c=0$.
As $\langle \bar{E}_{\nu_x}\rangle=\langle E_{\nu_x}\rangle$ in the neutrino
sphere and the propagating neutrinos are also in the background, in
Eq.~(\ref{neubmu}) we take $E_{\nu_x}=\langle \bar{E}_{\nu_x}\rangle=\langle E_{\nu_x}\rangle$.
Now the potential difference between $\nu_e$ and $\nu_{\mu}$ will be
\bary
V_{e\mu}&=&b_e-b_\mu=\sqrt2 G_F\biggl[(N_{\nu_e}-\bar{N}_{\nu_e})-(N_{\nu_\mu}-\bar{N}_{\nu_\mu})\nonumber\\
&&-\frac83\frac{1}{M_Z^2}\biggl\{\langle E_{\nu_e}\rangle\biggl(\langle E_{\nu_e}\rangle
N_{\nu_e}+\langle \bar{E}_{\nu_e}\rangle \bar{N}_{\nu_e}\biggr)
-\biggl(\langle E_{\nu_\mu}\rangle^2 N_{\nu_\mu}+\langle {E}_{\nu_\mu}\rangle^2 \bar{N}_{\nu_\mu}\biggr)\biggr\}\biggr].
\eary
Let us assume that the number density of neutrino and anti-neutrino of all flavors are the same inside the neutrino sphere, i.e.
\be
N_{\nu_l}=\bar{N}_{\nu_l}\,, \,\,\, \, l=e, \mu, \tau,
\label{neqnbar}
\ee
and this gives
\be
V_{e\mu}=2.91\times 10^{-18} N_{\nu_l}\, MeV^{-2}.
\ee
The potential difference between $\nu_{\mu}$ and $\nu_{\tau}$ vanishes
($V_{\mu\tau}=0$)
and the potential difference between $\nu_e$ and the sterile neutrino $\nu_s$ is given by
\be
V_{es}=b_e-b_s=-1.32\times 10^{-18} N_{\nu_l} \,MeV^{-2}.
\ee
But if we do not take into account the restriction given in Eq.~(\ref{neqnbar})
 then potential for $\nu_e$, $\nu_{\mu}$ and $\nu_{\tau}$ will different
from each other.


\section{GRB physics and Fireball Model}

We have already given a short introduction to GRB and fireball model in sec. 1.
As we are interested in the propagation of neutrinos in the fireball medium, let us
discuss a bit more about it. The fireball is formed due to the sudden release of huge
amount of energy in the form of gamma-rays into a compact region of size $c\delta t$  and
it will thermalize with a temperature around 3-10 MeV by producing electron, positron
pairs \cite{Zhang:2001ut}. It will also contaminated by baryons both from the progenitor and the
surrounding medium which is believed to be in the range
$10^{-8} M_{\odot}-10^{-5} M_{\odot}$.

Among the GRB community, it is strongly believed that the prompt $\gamma$ which we see
in the rage of few 100 keVs to few MeVs for few seconds is due to the synchrotron radiation
of charged particles in a magnetic field. But comparatively strong magnetic field is needed to
fit the observed data. But it is difficult to estimate the strength of the magnetic field from
the first principle. One would expect large magnetic field if the progenitors
are highly magnetized, for example, magnetars with $B\sim 10^{16}~G$. A
relatively small pre-existing magnetic field can be amplified due to turbulent
dynamo mechanism, compression or shearing. Also under suitable condition
the neutrino-electron interaction in the fireball plasma will be able to
amplify pre-existing small scale magnetic field. Despite all these, there is
no satisfactory explanation for the existence of strong magnetic field in the
fireball. Also even if some magnetic flux is carried by the outflow, it will
decrease due to the expansion of the fireball at a larger radius. But the strength
of the magnetic field will be much smaller than the critical field $B_C$. So the
derivation of the effective potential for weak field limit is justified here.
However, if we can measure the polarization of the GRBs, it will be helpful to estimate the
magnetic field
in the fireball as well as give information about the nature of the central engine.

Here we consider a CP-asymmetric $\gamma$ and $e^-e^+$ fireball,
where the excess of electrons come from the electrons associated
with the baryons within the fireball. Here for simplicity we assume that the
fireball is charge neutral $L_e=L_p$ and spherical
with an initial radius $R\simeq (100-1000)$ km and it has equal number of
protons and neutrons. Then the baryon load in the fireball can be given by
\bary
M_b &\simeq & \frac{16}{3\pi} \xi(3) L_e T^3 R^3 m_p\nonumber\\
    &\simeq & 2.23\times 10^{-4} L_e T^3_{MeV} R^3_7 M_{\odot}.
\eary
where $T_{MeV}$ is the fireball temperature expressed in MeV and lies in the range 3-10.
The quantity $R_7$ is in
units of $10^7$ cm and $m_p$ is the proton mass. For ultra relativistic
expansion of the fireball, we assume the baryon load in it to be in the range
$10^{-8}M_{\odot}-10^{-5}M_{\odot}$ which corresponds to lepton asymmetry
in the range $8.1\times 10^{-4} R^{-3}_7\le L_e \le 8.1\times 10^{-1} R^{-3}_7$.

We have already discussed about the different origins of 5-30 MeV neutrinos. Once
these neutrinos are produced, fractions of these neutrinos may propagate through
the fireball which is in an extreme condition and may affect the
propagation of these neutrinos through it.

\section{Three-Neutrino Mixing}

To find the neutrino oscillation probabilities, we have to solve the
Schroedinger's equation, given by
\be
i\frac{d\vec{\nu}}{dt}=H\vec{\nu},
\ee
and the state vector in the flavor basis is defined as
\be
\vec{\nu}\equiv(\nu_e,\nu_\mu,\nu_\tau)^T.
\ee
The effective Hamiltonian is
\be
H=U\cdot H^d_0\cdot U^\dagger+diag(V_e,0,0),
\ee
with
\be
H^d_0=\frac{1}{2E_\nu}diag(-\Delta m^2_{21},0,\Delta^2_{32}).
\ee
Here $V_e$ is the charge current (CC) matter potential and $U$ is the three
neutrino  mixing matrix given by\cite{GonzalezGarcia:2002dz,Akhmedov:2004ny}
\be
U =
{\pmatrix
{
c_{13}c_{12}                    & s_{12}c_{13}                    & s_{13}\cr
-s_{12}c_{23}-s_{23}s_{13}c_{12} & c_{23}c_{12}-s_{23}s_{13}s_{12}   & s_{23}c_{13}\cr
s_{23}s_{12}-s_{13}c_{23}c_{12}  &-s_{23}c_{12}-s_{13}s_{12}c_{23}   &  c_{23}c_{13}\cr
}},
\ee
where $s_{ij}=\sin\theta_{ij}$ and  $c_{ij}=\cos\theta_{ij}$. For
anti-neutrinos one has to replace $(N_a-{\bar N}_a)$ by $-(N_a-{\bar N}_a)$
and $U$ by $U^*$. The higher order contribution to potential does not change
the sign.
Here we have to emphasize that the neutral current (NC) contribution is not
taken into account. This is because in the matter background the NC
contribution to all the neutrinos is the same and when we take the difference
of potential, this contribution will be cancelled out and does not affect the
neutrino oscillation. But it has to be remembered that, in the neutrino
background where $N_{\nu}-{\bar N}_{\nu}\neq 0$, the potential for different
neutrinos are different which
described in Sec.\ref{neuback} and in this case we can not neglect the NC contribution.
The different neutrino probabilities are given as
\bary
P_{ee}&=&1-4s^2_{13,m}c^2_{13,m}S_{31},\nonumber\\
P_{\mu\mu}&=&1-4s^2_{13,m}c^2_{13,m}s^4_{23}S_{31}-4s^2_{13,m}s^2_{23}c^2_{23}S_{21}-4
c^2_{13,m}s^2_{23}c^2_{23}S_{32},\nonumber\\
P_{\tau\tau}&=&1-4s^2_{13,m}c^2_{13,m}c^4_{23}S_{31}-4s^2_{13,m}s^2_{23}c^2_{23}S_{21}-4
c^2_{13,m}s^2_{23}c^2_{23}S_{32},\nonumber\\
P_{e\mu}&=&4s^2_{13,m}c^2_{13,m}s^2_{23}S_{31},\nonumber\\
P_{e\tau}&=&4s^2_{13,m}c^2_{13,m}c^2_{23}S_{31}\nonumber\\
P_{\mu\tau}&=&-4s^2_{13,m}c^2_{13,m}s^2_{23}c^2_{23}S_{31}+4s^2_{13,m}s^2_{23}c^2_{23}S_{21}+4
c^2_{13,m}s^2_{23}c^2_{23}S_{32},\nonumber\\
\eary
where
\be
\sin
2\theta_{13,m}=\frac{\sin2\theta_{13}}{\sqrt{(\cos2\theta_{13}-2E_{\nu}V_e/\Delta
    m^2_{32})^2+(\sin2\theta_{13})^2}},
\ee
and
\be
S_{ij}=\sin^2\biggl(\frac{\Delta\mu^2_{ij}}{4E_{\nu}}L\biggr).
\ee
\bary
\Delta\mu^2_{21}&=&\frac{\Delta
  m^2_{32}}{2}\biggl(\frac{\sin2\theta_{13}}{\sin2\theta_{13,m}}-1\biggr)-E_{\nu}V_e\nonumber\\
\Delta\mu^2_{32}&=&\frac{\Delta
  m^2_{32}}{2}\biggl(\frac{\sin2\theta_{13}}{\sin2\theta_{13,m}}+1\biggr)+E_{\nu}V_e\nonumber\\
\Delta\mu^2_{31}&=&\Delta m^2_{32} \frac{\sin2\theta_{13}}{\sin2\theta_{13,m}}
\eary

where
\bary
\sin^2\theta_{13,m}&=&\frac12\biggl(1-\sqrt{1-\sin^22\theta_{13,m}}\biggr)\nonumber\\
\cos^2\theta_{13,m}&=&\frac12\biggl(1+\sqrt{1-\sin^22\theta_{13,m}}\biggr)
\eary
The oscillation length for the neutrino is given by
\be
L_{osc}=\frac{L_v}{\sqrt{\cos^2 2\theta_{13} (1-\frac{2 E_{\nu} V_e}{\Delta m^2_{32} \cos 2\theta_{13}}
    )^2+\sin^2 2\theta_{13}}},
\label{osclength}
\ee
where $L_v=4\pi E_{\nu}/\Delta m^2_{32}$ is the vacuum oscillation length.
For resonance to occur, we should have $V_{eff,B}=V_e>0$ and
\be
\cos 2\theta_{13} = \frac{2 E_{\nu} V_e}{ \Delta m^2_{32}}.
\label{reso}
\ee
By putting $V_e$ and simplifying we obtain
\be
\Phi_A-1.58027\times 10^{-10} E_{\nu,MeV} \Phi_B \simeq 2.24208 \frac{\tilde
  \Delta m^2_{32}}{E_{\nu, MeV}} \cos 2\theta_{13},
\label{resocond}
\ee
%
\begin{table}
\caption{\small\sf We have shown the different observables of the
fireball as well as the neutrino resonance length $L_{res}$ for
$\Delta m^2_{23}=10^{-2.9}eV^2$ and $B=0.1 B_c$. }
\begin{center}\renewcommand{\tabcolsep}{0.38cm}
\renewcommand{\arraystretch}{1.05}
\begin{tabular}{|c|c|c|c|c|c|}\hline
$E_{\nu,MeV}$&T(MeV)&$\mu(eV)$&$L_e$&$L_{res}(cm)$& $M_b(R^3_7 M_\odot)$\\ \hline
5&3&$2.47412$&$1.099\times 10^{-6}$  &$4.737\times 10^6 $&$ 7.064\times 10^{-9}$\\
 &10&$0.960329$&$1.284\times 10^{-7}$&&$3.057\times 10^{-8}$\\\hline
10&3& $1.33789$  &$5.943\times 10^{-7}$   &$9.474\times 10^6$ &$3.819\times 10^{-9}$ \\
  &10& $1.59669 $ &  $ 2.135\times 10^{-7}$ &&$ 5.083\times 10^{-8}$       \\ \hline
20&3 &$0.869801$& $ 3.864\times 10^{-7}$ &$1.895\times 10^7$ &$ 2.483\times 10^{-9}$  \\
  &10&$3.03194$ &$ 4.055\times 10^{-7}$ &&$ 9.653\times 10^{-8}$       \\ \hline
30&3& $0.804489$  &$ 3.574\times 10^{-7}$ &$2.842\times 10^7$ &$ 2.297\times 10^{-9}$        \\
  &10&$4.50277$ &$ 6.022\times 10^{-7}$ &&$ 1.434\times 10^{-7}$        \\ \hline
\end{tabular}
\label{}
\end{center}
\end{table}
\begin{table}
\caption{\small\sf We have shown the different observables of the
fireball as well as the neutrino resonance length $L_{res}$ for
$\Delta m^2_{23}=10^{-2.2}eV^2$ and $B=0.1 B_c$. }
\begin{center}\renewcommand{\tabcolsep}{0.38cm}
\renewcommand{\arraystretch}{1.05}
\begin{tabular}{|c|c|c|c|c|c|}\hline
$E_{\nu,MeV}$   & T(MeV)&$\mu(eV)$& $L_e$ & $L_{res}(cm)$  & $M_b (R^3_7 M_\odot)$  \\ \hline
5&3&$12.1177$   &$5.383\times 10^{-6}$&$9.452\times 10^5$  &$3.460\times 10^{-8}$        \\
 &10& $1.82987$   & $ 2.447\times 10^{-7}$ &&$5.826\times 10^{-8}$    \\\hline
10&3&    $6.17192$ & $ 2.742\times 10^{-6}$&$1.890\times 10^6$ &$1.762\times 10^{-8}$  \\
&10 &  $2.03432$ &    $ 2.721\times 10^{-7}$ &&$6.477\times 10^{-8}$    \\ \hline
20& 3&$3. 27657$& $ 1.456\times 10^{-6}$&$3.780\times 10^6$&$9.355\times 10^{-9}$       \\
& 10& $3.24878$ & $ 4.346\times 10^{-7}$  &&$1.034\times 10^{-7}$     \\ \hline
30& 3&$2.4178$ &  $ 1.074\times 10^{-6}$ &$5.671\times 10^6$&$6.903\times 10^{-9}$      \\
 & 10& $4.6475$& $ 6.216\times 10^{-7}$ & &$1.480\times 10^{-7}$     \\ \hline
\end{tabular}
\label{}
\end{center}
\end{table}
\begin{figure}[t!] 
\vspace{0.5cm}
{\centering
\resizebox*{0.4\textwidth}{0.3\textheight}
{\includegraphics{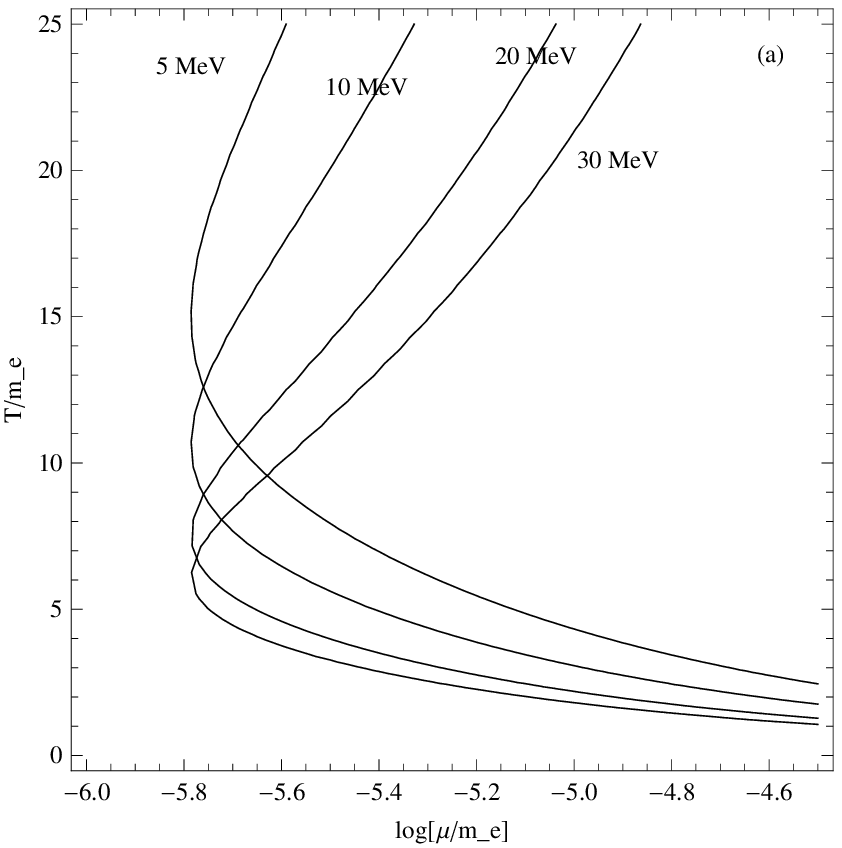}}
\resizebox*{0.4\textwidth}{0.3\textheight}
{\includegraphics{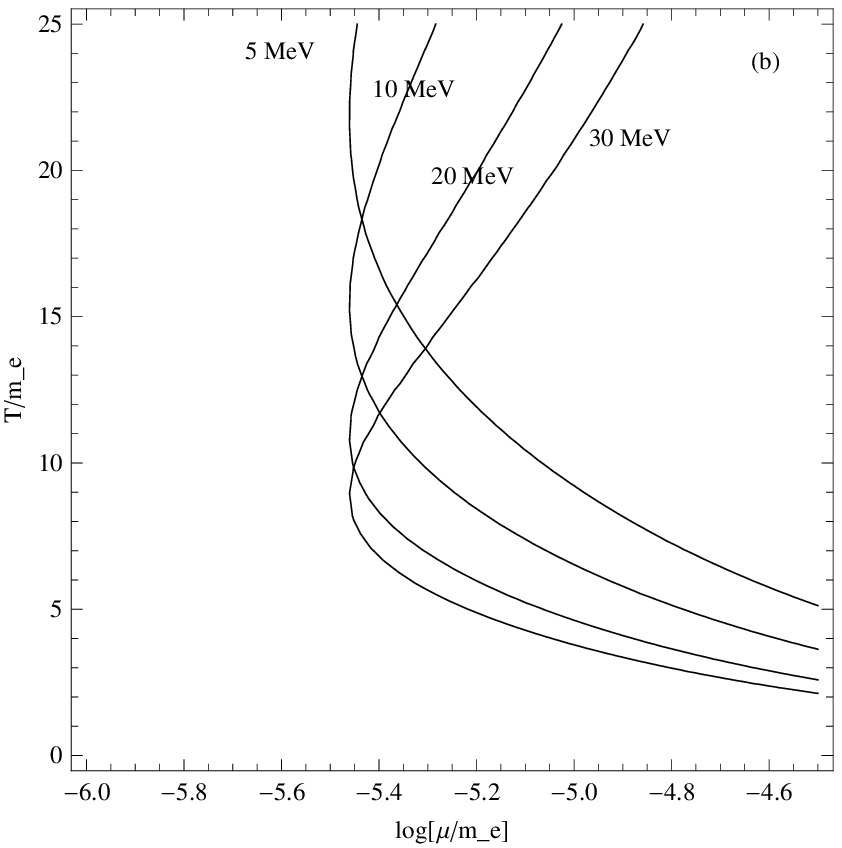}}
}
\caption{\small\sf
The contour plot of the resonance condition of Eq.~(\ref{resocond}) as
functions of $T/m_e$ and $\log\left [\frac{\mu}{m_e}\right ]$ is shown for different
neutrino energies and $B=0.1 B_c$ where (a) is for
$\Delta m^2_{32}=10^{-2.9}\, eV^2$ and  (b) is for
$\Delta m^2_{32}=10^{-2.2}\, eV^2$.  }
\label{figure3}
\end{figure}
where ${\tilde\Delta m^2_{32}}$ is expressed in $eV^2$ and $E_{\nu,
  MeV}$ is in MeV. The functions $\Phi_A$ and $\Phi_B$ are defined in Eqs.~(\ref{fphiA}) and
(\ref{fphiB}). Now we have to evaluate the above condition for given values of
${\tilde\Delta m^2_{32}}$ and $\cos 2\theta_{13}$ from experiments and different
values of temperature (T) and chemical potential ($\mu$). At resonance, the oscillation
length becomes the resonance length and can be given by
\be
L_{res}=\frac{L_v}{\sin 2\theta_{13}}.
\ee
So far we have assumed that the neutrino potential does not vary with distance.
However $V_{eff,B}$ will vary with distance. So we have to consider the adiabatic
condition at the resonance, which can be given by
\bary
\kappa_{res}&\equiv & \frac{2}{\pi}
\left ( \frac{\Delta m^2_{32}}{2 E_{\nu}} \sin 2\theta_{13}\right )^2
\left (\frac{dV_{eff,B}}{dr}\right)^{-1} \ge 1\nonumber\\
&=& 3.62\times 10^{-2}
\left ( \frac{{\tilde \Delta m}^2_{32}}{E_{\nu,MeV}} \sin 2\theta_{13}\right )^2
\frac{l_{cm}}{\Phi^{\prime}}\ge 1,
\label{adbcon}
\eary
where
\be
\Phi^{\prime}=\frac{d\Phi_A}{dx}
- 1.58027\times 10^{-10} E_{\nu,MeV} \frac{d\Phi_B}{dx}.
\ee
In the above equations we have expressed $l_{cm}$ in centimeter and $x$
is a dimensionless variable.


\begin{figure}[t!] 
\vspace{0.5cm}
{\centering
\resizebox*{0.4\textwidth}{0.3\textheight}
{\includegraphics{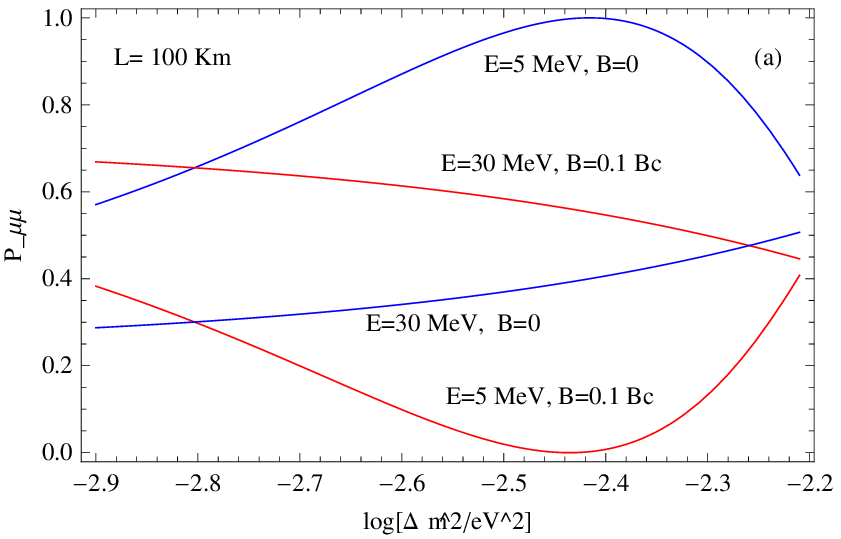}}
\resizebox*{0.4\textwidth}{0.3\textheight}
{\includegraphics{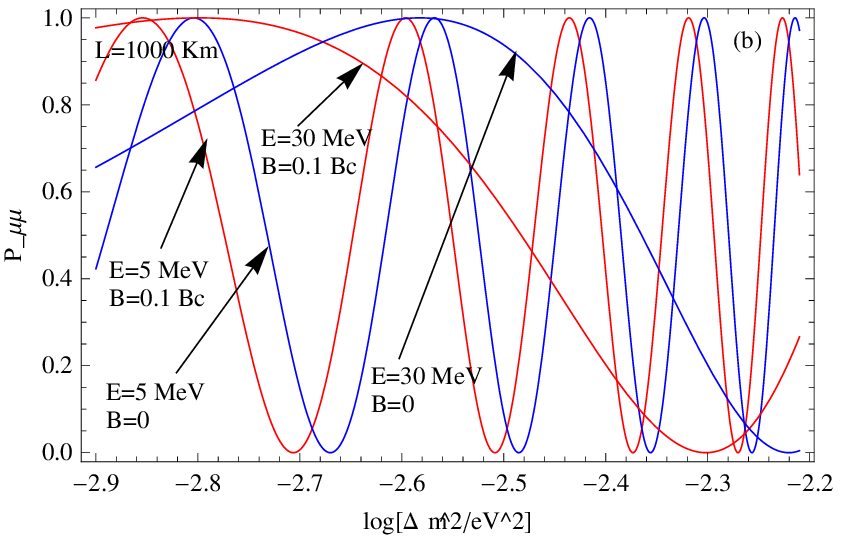}}
}
\caption{\small\sf The survival probability of muon neutrinos $P_{\mu\mu}$
is plotted as a function of $\log \left[\frac{\Delta m^2_{32}}{eV^2}\right ]$, for the
fireball radius $L=100\,km$ (a) and $L=1000\,km$ (b). The neutrino energy
and magnetic field are shown in it.}
\label{figure4}
\end{figure}
\begin{figure}[t!] 
\vspace{0.5cm}
{\centering
\resizebox*{0.4\textwidth}{0.3\textheight}
{\includegraphics{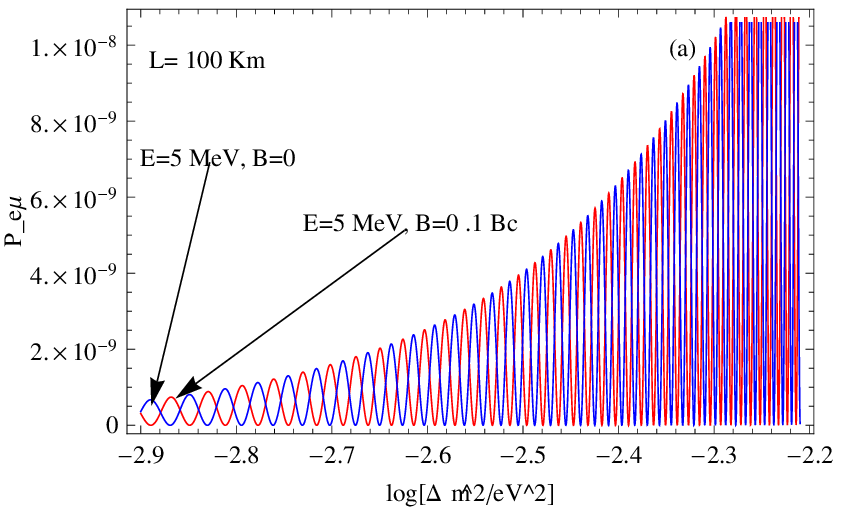}}
\resizebox*{0.4\textwidth}{0.3\textheight}
{\includegraphics{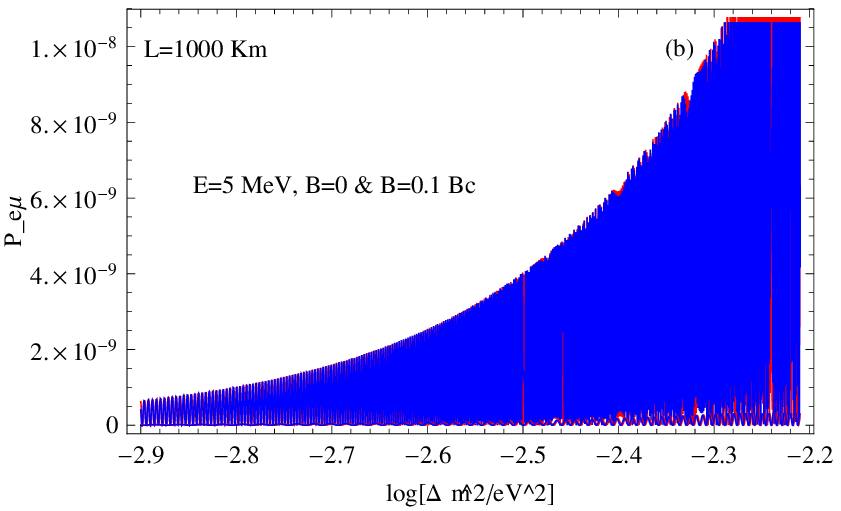}}
}
\caption{\small\sf The probability $P_{e\mu}$ is plotted as a function of
$\log \left [\frac{\Delta m^2_{32}}{eV^2}\right ]$, for the fireball radius $L=100\,km$ (a)
and $L=1000\,km$ (b) . The neutrino energy
and magnetic field are shown in it.}
\label{figure5}
\end{figure}
\begin{figure}[t!] 
\vspace{0.5cm}
{\centering
\resizebox*{0.4\textwidth}{0.3\textheight}
{\includegraphics{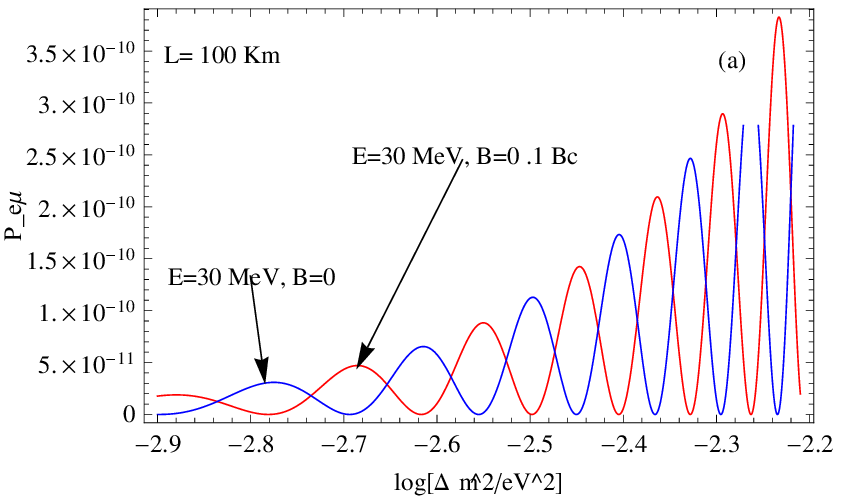}}
\resizebox*{0.4\textwidth}{0.3\textheight}
{\includegraphics{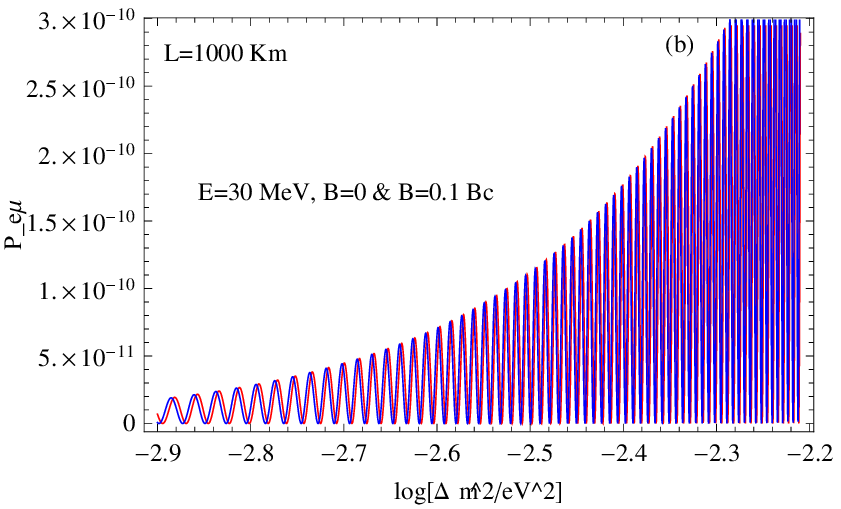}}
}
\caption{\small\sf This is same as Fig. 5 but for neutrino energy
$E_{\nu}=30\, MeV$}
\label{figure6}
\end{figure}
\begin{figure}[t!] 
\vspace{0.5cm}
{\centering
\resizebox*{0.4\textwidth}{0.3\textheight}
{\includegraphics{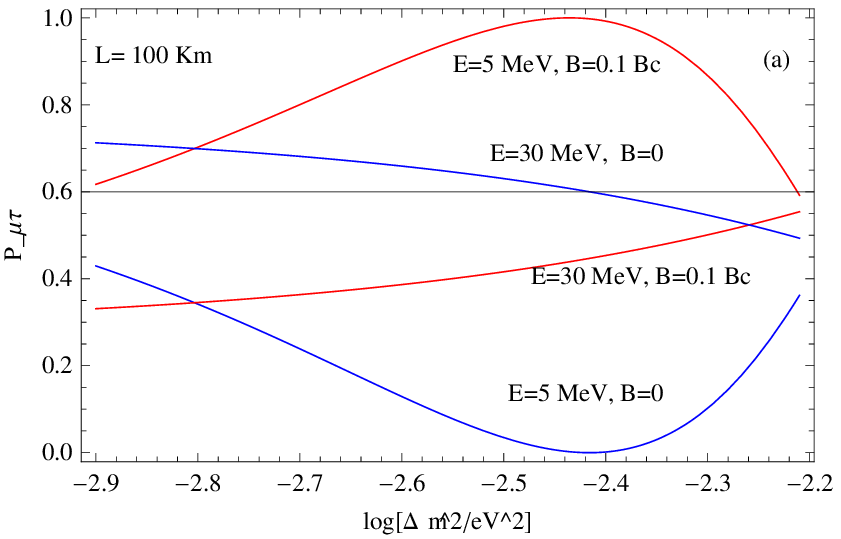}}
\resizebox*{0.4\textwidth}{0.3\textheight}
{\includegraphics{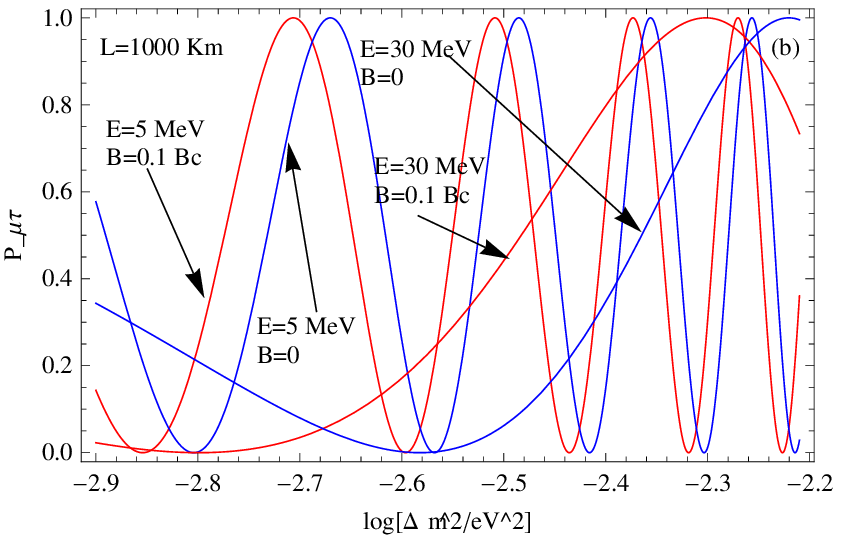}}
}
\caption{\small\sf The probability $P_{\mu\tau}$
is plotted as a function of $\log \left [\frac{\Delta m^2_{32}}{eV^2}\right ]$, for the
fireball radius $L=100\,km$ (a) and $L=1000\,km$ (b) . The neutrino energy
and magnetic field are shown in it.}
\label{figure7}
\end{figure}

\section{Results}

We have done a complete analysis for three neutrino mixing with and without
magnetic field. For our analysis we have used the result
given in ref.\cite{GonzalezGarcia:2002dz},
\bary
&&1.4\times 10^{-3} < \Delta m^2_{32}/eV^2 < 6.0\times 10^{-3}\nonumber\\
&&\theta_{13} \simeq 6^{\circ}\nonumber\\
&&32^{\circ} < \theta_{23} < 60^{\circ}.
\eary
The above result is obtained by performing a global analysis and taking
full set of data from solar, atmospheric and reactor experiments. In the above
we consider $\theta_{23}=45^{\circ}$.

Different values of $\mu$ and $T$ are shown in Fig. 3  for which the
resonance condition in Eq.~(\ref{resocond}) is satisfied. We have used
two extreme values of $\Delta m^2_{32}$ i.e.
$10^{-2.9}\,eV^2$ in Fig. 3a and  $10^{-2.21}\,eV^2$ in Fig. 3b for $B=0.1 B_c$,
$\theta_{13}=6^{\circ}$ and four different neutrino energies $5$, $10$, $20$
and $30$ MeV respectively. In both small and large values of $\Delta m^2_{32}$
and fireball temperature in the range $3$ to $10$ MeV, the chemical potential
of the electron is in the 1-12 eV range which are shown in Table I and II.
In Table I we have shown the different fireball observables for
$\Delta m^2_{32}=10^{-2.9}\,eV^2$. It shows that going from neutrino energy
5 MeV to 30 MeV, the resonance length changes between 47 km to 284 km and
lepton asymmetry $~\sim 10^{-7}$ to $~10^{-6}$. For a charge neutral plasma
$L_e=L_p$ this translates to a baryon load in the fireball
in the range $2.3\times 10^{-9}M_{\odot} < M_b < 1.4\times 10^{-7}M_{\odot}$.
In Table II we have done the same analysis but for
$\Delta m^2_{32}=10^{-2.2}\,eV^2$. Here the chemical potential is higher
compared to the one in Table I. This shows that shift in $\Delta m^2_{32}$
towards higher value also shift the $\mu$ in the same direction. In this case
there is not much change in $L_e$ and $M_b$ and the resonance length lies in
the range $9.4\,km <L_{res}< 57\,km$.

The above analysis of the resonance condition shows that, in the temperature
range of 3 to 10 MeV, the resonance condition is satisfied for electron
chemical potential ($\mu$) between 1 to 12 eV. Also for neutrino energy in
the range 5 to 30 MeV, the resonance length lies below 284 km, which shows
that neutrinos can resonantly oscillate within the fireball of radius 100 to
1000 km. The baryon load of the fireball also lies in the range
$10^{-9} M_{\odot} < M_b < 10^{-7} M_{\odot}$.
If the potential has a profile that means, the
functions $\Phi_A$ and $\Phi_B$ depend on the length scale then we found that
for $ 10^{-2.9}\,eV^2 \le \Delta m^2_{32} \le 10^{-2.2}\,eV^2$ and $5\le E_{\nu, MeV}\le 30$,
the parameter $l_{cm}/{\Phi^{\prime}}$ will lie in the range $10^{-10}$ to
$2.5\times 10^{-9}$ to satisfy the condition given in Eq.~(\ref{adbcon}).

The survival and conversion probabilities for the active neutrinos are
plotted as function of $\Delta m^2_{32}$ in the range $10^{-2.9}~eV^2$
to $10^{-2.2}~eV^2$
for $B=0$ and $B=0.1 B_c$ for two neutrino energies 5 MeV and 30 MeV from
Fig. 4 to Fig. 7. We have also considered two different length scales for the
fireball i.e. 100 km and 1000 km to see how the probabilities changes when the
length scale of the fireball changes. As we are taking $\theta_{23}=45^{\circ}$,
the probabilities $P_{e\mu}=P_{e\tau}$ and also $P_{\mu\mu}=P_{\tau\tau}$.

We have plotted the
survival probability of muon neutrino for $L=100\,km$ in Fig. 4a and
for  $L=1000\,km$ in Fig. 4b. The survival probability of muon $P_{\mu\mu}$
neutrino in
Fig. 3a, for neutrino energy 5 MeV and magnetic field $B=0$ is $180^{\circ}$
out of phase compared to the same neutrino energy but for $B=0.1 B_c$. For
$B=0$ case the probability varies between $0.6$ and unity and for $B=0.1 B_c$
it is between $0$ and $0.4$. Going from 5 MeV to 30 MeV we saw that,
for $B=0$,the $P_{\mu\mu}$ decreases and lies between $0.3$ and $0.5$ and for
$B=0.1 B_c$ lies between $0.42$ and $0.68$. Going from $L=100\,km$ to
$L=1000\, km$ (Fig. 4b), we see that both $B=0$ and $B=0.1 B_c$ have a small
phase difference and the $P_{\mu\mu}$ varies between $0$ and $1$. But
$P_{\mu\mu}$ for $B=0.1 B_c$ lags behind the one for $B=0$ for both neutrino
energies 5 MeV and 30 MeV. The probability for low energy (5 MeV curve) neutrino
oscillates faster than the one for 30 MeV.

In Fig. 5, we have plotted the conversion probability $P_{e\mu}$
for $E_{\nu}=5\,MeV$, $L=100\,km$ (5a) and $E_{\nu}=5\,MeV$, $L=1000\,km$ (5b)
respectively for both $B=0$ and $B=0.1 B_c$. It shows that in Fig. 5a,
the $P_{e\mu}$ for $B=0$ and $B=0.1 B_c$
are having the same phase difference of $180^{\circ}$ as in Fig. 4a.
But the probability is very small $\sim 10^{-9}$. Going from Fig. 5a to 5b
($L=1000\,km$) , we saw that the phase difference is almost gone away and
$P_{e\mu}$ oscillates much faster compared to the one in Fig 5a.

In Figs. 6a and 6b, we have the same  probability $P_{e\mu}$ as in Fig. 5,
but here the neutrino energy $E_{\nu}=30\,MeV$. It is clearly seen in
Fig. 6a that both the probabilities are out of phase and are very small $\sim 10^{-10}$.
On the other hand for $L=1000\,km$ (Fig. 6b) the phase difference is gone and
the $P_{e\mu}$ oscillates much faster than the one in Fig. 6a. We have the
$P_{e\mu}=P_{e\tau}\simeq 10^{-10}$ which gives $P_{ee}\simeq 1$. This shows that
the electron neutrinos propagating within the fireball can not oscillate to
other neutrinos.

In Fig. 7a and 7b we have plotted the $P_{\mu\tau}$ for $L=100\,km$ and
$L=1000\,km$ respectively. For $L=100\,km$ the probability for $B=0$ and
$B=0.1 B_c$ are out of phase for both $E_{\nu}=5$ and $30 MeV$ in Fig 7a. In
Fig. 7b for $L=1000\,km$ there is a small phase difference between the $B=0$
and $B=0.1 B_c$ probabilities. Comparison of Fig. 4a with Fig. 7a and Fig. 4b with Fig. 7b show
that the $B=0$ probability (in Fig. 4a and Fig. 7a) and $B=0.1 B_c$ probability
(in Fig. 4b and Fig. 7b) are $180^{\circ}$ out of phase. We obtain this because
the probability satisfies the condition
\be
P_{\mu\mu}+P_{e\mu}+P_{\mu\tau}=1.
\ee
We have shown in Figs. 5 and 6 that $P_{e\mu}=P_{e\tau}$ and they are very small which gives
$P_{\mu\tau}\simeq 1-P_{\mu\mu}$. The $P_{\tau\tau}$ is same as $P_{\mu\mu}$.

From our analysis we see that $P_{ee}\simeq 1$ and is almost independent of the
energy of the neutrinos and the size of the fireball, which shows that
for small mixing corresponding to $\theta_{13}=6^{\circ}$ and for
$\theta_{23}=45^{\circ}$, the electron neutrino almost does not oscillate to any
other flavor which is obvious from the Fig.~\ref{figure5} and
Fig.~\ref{figure6}.
On the other hand, the muon and tau neutrinos oscillate among themselves with
equal probability and the oscillation depends on the neutrino energy, magnetic
field and size of the fireball.
Comparison of $B=0$ and $B\neq 0$ results show that the magnetic field
contribution is order of magnitude smaller than the medium case. But depending
on the size of the fireball, the probability for $B=0$ and $B\neq 0$ are either
in phase or out of phase.

\section{conclusions}

We have shown that neutrino self-energy in the presence of a magnetic field
can also be expressed as
\be
{\tilde \Sigma}=a_{\perp}\rlap /k_{\perp}+b\rlap /u+c\rlap /b,\nonumber\\
\label{sigmanb3}
\ee
by absorbing the ${\rlap /k}_{\parallel}$ component with the two four vectors
${\rlap /u}$ and ${\rlap /b}$. The above decomposition is only
valid when the magnetic field is along the z-axis.
In addition to this we have also shown that
the neutrino
effective potential $V_{eff,B}$ is independent of how we decompose the
the self-energy in terms of Lorentz scalars as shown in Eqs.~(\ref{sigmanb}) and
(\ref{sigmanb3}).
We have explicitly calculated the $V_{eff,B}$ up to order
$M^{-4}_W$ in the weak field limit $eB\ll m^2_e$ in terms of Bessel Functions
and recover the result for $B=0$ limit which can only be obtained when
$k_3$ the third component of the neutrino momentum is replaced by
$-E_{\nu}$. We have also calculated the
neutrino effective potential when the background contains (i) $e^-,e^+$,
protons, neutrons and neutrinos and (ii) only neutrinos in the background.
By considering the three-neutrino mixing we have studied the
active-active neutrino oscillation process
$\nu_a\leftrightarrow \nu_b$ ($a$ and $b$ are active) in the weakly
magnetized  $e^-e^+$, $p$ and $n$ plasma of the GRB fireball
assuming it to be spherical with a radius of 100 to 1000 km
and temperature in the range 3-10 MeV. We further assume that the fireball is
charge neutral due to the presence of protons and their accompanying
electrons. The baryon load of the fireball is solely due to the presence of
almost equal number of protons and neutrons in it.

Our analysis shows that the
$\nu_e$ almost does not oscillate to any other flavors and $P_{ee}\simeq 1$ is independent of the
$\nu_e$ energy as well as the background magnetic field.
The non oscillation of $\nu_e$ to other flavors gives
$P_{e\mu}$ and $P_{e\tau}$ very small
of the order of $10^{-10}$. But the $\nu_{\mu}$ and $\nu_{\tau}$ oscillate
among themselves, which depends on the energy of the neutrino, magnetic field
and also on the size of the fireball. We analyzed our result by taking two
different radius of the fireball i.e. 100 km and 1000 km, neutrino energy
in the range 5 to 30 MeV and magnetic field $B=0$ and $B=0.1 B_c$ with $B_c=4.14 \times 10^{13}$ Gauss. 
We found that the probability for $B=0$ and $B=0.1 B_c$ are out of phase by
$180^{\circ}$ for $L=100$ km and almost in phase for $L=1000$ km. For
$L=100$ km, the $P_{\mu\mu}$ and  $P_{\mu\tau}$ do vary between 0-0.5 or 0.5-1.
On the other hand for $L=1000$ km, it varies between 0 and 1. Also in this case
low energy neutrinos oscillate faster than the high energy one.

We have also analyzed the resonance condition and found that, to satisfy the
resonance condition, the electron chemical potential in the fireball
lies in the range 1-12 eV. For neutrino energy in the range 5 to 30 MeV
the resonance length lies in the range 9.4 km to 284 km. So if we consider
a fireball of 100 km to 1000 km radius this shows that the resonant
oscillation of $\nu_{\mu}$ and $\nu_{\tau}$ neutrinos can take place but
not the $\nu_e$. The baryon load calculated by using the resonance condition
lies in the range $10^{-9} M_{\odot} < M_b < 10^{-7} M_{\odot}$.
Depending on the size of the fireball the probability for $B=0$ and $B\neq 0$
are either in phase or out of phase.

\vskip2.0cm
{\bf ACKNOWLEDGMENTS}\\
We are thankful to B. Zhang and S. Nagataki for many useful discussions.
Y.Y.~K and S.~S. thank APCTP  for the  kind hospitality during their several visits,
where this work has been initiated.
The Work of S.~S. is partially supported by  DGAPA-UNAM (Mexico) project
IN101409,
Y.Y.K's work is partially supported by APCTP in Korea and
is supported in part by National Research Foundation of Korea Grant funded by the Korean Government 2009-0070667.


\begin{thebibliography}{100}
\expandafter\ifx\csname natexlab\endcsname\relax\def\natexlab#1{#1}\fi
\expandafter\ifx\csname bibnamefont\endcsname\relax
  \def\bibnamefont#1{#1}\fi
\expandafter\ifx\csname bibfnamefont\endcsname\relax
  \def\bibfnamefont#1{#1}\fi
\expandafter\ifx\csname citenamefont\endcsname\relax
  \def\citenamefont#1{#1}\fi
\expandafter\ifx\csname url\endcsname\relax
  \def\url#1{\texttt{#1}}\fi
\expandafter\ifx\csname urlprefix\endcsname\relax\def\urlprefix{URL }\fi
\providecommand{\bibinfo}[2]{#2}
\providecommand{\eprint}[2][]{\url{#2}}



\bibitem{Elmfors:1996gy}
  P.~Elmfors, D.~Grasso and G.~Raffelt,
  Nucl.\ Phys.\  B {\bf 479}, 3 (1996)
  [arXiv:hep-ph/9605250].
		

\bibitem{Koers:2005ya}
  H.~B.~J.~Koers and R.~A.~M.~Wijers,
  Mon.\ Not.\ Roy.\ Astron.\ Soc.\  {\bf 364}, 934 (2005)
  [arXiv:astro-ph/0505533].

\bibitem{Dessart:2008zd}
  L.~Dessart, C.~Ott, A.~Burrows, S.~Rosswog and E.~Livne,
  arXiv:0806.4380 [astro-ph].


\bibitem{Sahu:2009iy}
  S.~Sahu, N.~Fraija and Y.~Y.~Keum,
  Phys.\ Rev.\  D {\bf 80}, 033009 (2009)
  [arXiv:0904.0138 [hep-ph]].

\bibitem{Langacker:1982ih}
  P.~Langacker, J.~P.~Leveille and J.~Sheiman,
Phys.\ Rev.\  D {\bf 27}, 1228 (1983).

\bibitem{BravoGarcia:2007uc}
  A.~Bravo Garcia, K.~Bhattacharya and S.~Sahu,
  Mod.\ Phys.\ Lett.\  A {\bf 23}, 2771 (2008)
  [arXiv:0706.3921 [hep-ph]].




\bibitem{Piran:1999kx}
  T.~Piran,
  Phys.\ Rept.\  {\bf 314} (1999) 575
  [arXiv:astro-ph/9810256].

\bibitem{Meegan:1992xg}
  C.~A.~Meegan {\it et al.},
  Nature {\bf 355}, 143 (1992).



\bibitem{Zhang:2003uk}
  B.~Zhang and P.~Meszaros,
  Int.\ J.\ Mod.\ Phys.\  A {\bf 19}, 2385 (2004)
  [arXiv:astro-ph/0311321].

\bibitem{Piran:1999bk}
  T.~Piran,
  Phys.\ Rept.\  {\bf 333}, 529 (2000)
  [arXiv:astro-ph/9907392].

\bibitem{DellaValle:2005cr}
  M.~Della Valle,
  Nuovo Cim.\  {\bf 28C}, 563 (2005)
  [arXiv:astro-ph/0504517].


\bibitem{Gehrels:2005qk}
  N.~Gehrels {\it et al.},
  Nature {\bf 437}, 851 (2005)
  [arXiv:astro-ph/0505630].

\bibitem{Barthelmy:2005bx}
  S.~D.~Barthelmy {\it et al.},
  Nature {\bf 438}, 994 (2005)
  [arXiv:astro-ph/0511579].

\bibitem{Villasenor:2005xj}
  J.~S.~Villasenor {\it et al.},
 Nature {\bf 437}, 855 (2005)
  [arXiv:astro-ph/0510190].

\bibitem{Hjorth:2005}
  J.~Hjorth {\it et al.},
 Nature {\bf 437}, 859 (2005)
  [arXiv:astro-ph/0510096].

\bibitem{Berger:2005}
  E.~Berger {\it et al.},
 Nature {\bf 438}, 988 (2005)
  [arXiv:astro-ph/0508115].

\bibitem{Usov:1992zd}
  V.~V.~Usov,
  Nature {\bf 357}, 472 (1992).

\bibitem{Uzdensky:2007uf}
  D.~A.~Uzdensky and A.~I.~MacFadyen,
  Phys.\ Plasmas {\bf 14}, 056506 (2007)
  [arXiv:0707.0576 [astro-ph]].
  
\bibitem{uffo:2009}
I.H.~Park et al.(UFFO collaboration),
The UFFO(Ultra-Fast Flash Observatory) Pathfinder: Proposed Space Mission for Lomonosov Spacecraft.  

\bibitem{Waxman:2003vh}
  E.~Waxman,
  Lect.\ Notes Phys.\  {\bf 598} (2003) 393
  [arXiv:astro-ph/0303517].


\bibitem{Goodman:1986az}
  J.~Goodman,
  Astrophys.\ J.\  {\bf 308} (1986) L47.


\bibitem{Vedrenne}
  G.~Vedrenne and J.~Atteia,
 ``Gamma-Ray Bursts: The brighest Explotions in the Universe,''
  Springer, Praxis Publishing Ltd, Chichester, UK, 2009

\bibitem{Lee:2007js}
  W.~H.~Lee and E.~Ramirez-Ruiz,
  New J.\ Phys.\  {\bf 9}, 17 (2007)
  [arXiv:astro-ph/0701874].


\bibitem{Raffelt:2001kv}
  G.~G.~Raffelt,
  Astrophys.\ J.\  {\bf 561}, 890 (2001)
  [arXiv:astro-ph/0105250].


\bibitem{Ruffert:1998qg}
  M.~Ruffert and H.~T.~Janka,
  Astron.\ Astrophys.\  {\bf 344}, 573 (1999)
  [arXiv:astro-ph/9809280].

\bibitem{Goodman:1986we}
  J.~Goodman, A.~Dar and S.~Nussinov,
 Astrophys.\ J.\  {\bf 314} (1987) L7.

\bibitem{Waxman:1997ti}
  E.~Waxman and J.~N.~Bahcall,
  Phys.\ Rev.\ Lett.\  {\bf 78}, 2292 (1997)
  [arXiv:astro-ph/9701231].

\bibitem{Volkas:1999gb}
  R.~R.~Volkas and Y.~Y.~Y.~Wong,
  Astropart.\ Phys.\  {\bf 13}, 21 (2000)
  [arXiv:astro-ph/9907161].

\bibitem{Dasgupta:2008cu}
  B.~Dasgupta, A.~Dighe, A.~Mirizzi and G.~G.~Raffelt,
  Phys.\ Rev.\  D {\bf 78}, 033014 (2008)
  [arXiv:0805.3300 [hep-ph]].



\bibitem{Nunokawa:1997dp}
  H.~Nunokawa, V.~B.~Semikoz, A.~Y.~Smirnov and J.~W.~F.~Valle,
[arXiv:hep-ph/9701420].





\bibitem{Nieves:1990ne}
  J.~F.~Nieves,
  Phys.\ Rev.\  D {\bf 42}, 4123 (1990)
  [Erratum-ibid.\  D {\bf 49}, 3067 (1994)].

\bibitem{Weldon:1982aq}
  H.~A.~Weldon,
  Phys.\ Rev.\  D {\bf 26}, 1394 (1982).


\bibitem{D'Olivo:2002sp}
  J.~C.~D'Olivo, J.~F.~Nieves and S.~Sahu,
  Phys.\ Rev.\  D {\bf 67}, 025018 (2003)
  [arXiv:hep-ph/0208146].

\bibitem{Erdas:1990gy}
  A.~Erdas and G.~Feldman,
  Nucl.\ Phys.\  B {\bf 343}, 597 (1990).


\bibitem{Schwinger}
J. Schwinger
Phys.\ Rev.\ Lett.{\bf 82}, 5, 664, (1951)



\bibitem{Enqvist:1990ad}
  K.~Enqvist, K.~Kainulainen and J.~Maalampi,
  Nucl.\ Phys.\  B {\bf 349}, 754 (1991).

\bibitem{Sahu:1998jh}
  S.~Sahu and V.~M.~Bannur,
  Phys.\ Rev.\  D {\bf 61}, 023003 (2000)
  [arXiv:hep-ph/9806427].

\bibitem{Sahu:2005zh}
  S.~Sahu and J.~C.~D'Olivo,
  Phys.\ Rev.\  D {\bf 71}, 047303 (2005)
  [arXiv:hep-ph/0502043].

\bibitem{Erdas:1998uu}
  A.~Erdas, C.~W.~Kim and T.~H.~Lee,
  Phys.\ Rev.\  D {\bf 58}, 085016 (1998)
  [arXiv:hep-ph/9804318].



\bibitem{Garcia:2007ij}
  A.~Bravo.~Garcia and S.~Sahu,
  Mod.\ Phys.\ Lett.\  A {\bf 22} (2007) 213.


\bibitem{Giunti}
C.~Giunti and C.~W.~Kim,
Fundamentals of Neutrino Physics and Astrophysics, Oxford Univ.Press, pg 525 (2007)

\bibitem{Meszaros:1999fr}
  P.~Meszaros,
  Nucl.\ Phys.\ Proc.\ Suppl.\  {\bf 80}, 63 (2000)
  [arXiv:astro-ph/9904038].

\bibitem{GonzalezGarcia:2002dz}
  M.~C.~Gonzalez-Garcia and Y.~Nir,
  Rev.\ Mod.\ Phys.\  {\bf 75}, 345 (2003)
  [arXiv:hep-ph/0202058].

\bibitem{Akhmedov:2004ny}
  E.~K.~Akhmedov, R.~Johansson, M.~Lindner, T.~Ohlsson and T.~Schwetz,
  JHEP {\bf 0404}, 078 (2004)
  [arXiv:hep-ph/0402175].


\bibitem{Zhang:2001ut}
  B.~Zhang and P.~Meszaros,
  Astrophys.\ J.\  {\bf 566}, 712 (2002)
  [arXiv:astro-ph/0108402].


\end{thebibliography}
\end{document}